\documentclass[preprint,12pt]{emulateapj}

\usepackage{natbib}
\usepackage{verbatim}
\usepackage{graphicx,epstopdf}
\DeclareGraphicsExtensions{.ps}
\usepackage[space]{grffile}
\usepackage{latexsym}
\usepackage{amsfonts,amsmath,amssymb}
\usepackage{url}
\usepackage[utf8]{inputenc}
\usepackage{fancyref}
\usepackage{hyperref}
\hypersetup{colorlinks=false,pdfborder={0 0 0},}
\usepackage{textcomp}
\usepackage{longtable}
\usepackage{multirow,booktabs}
\usepackage[caption=false]{subfig}
\def\icarus{{Icarus}}
\newcommand{\ktwo}{{\it K2}\xspace}

\usepackage{xspace}
\usepackage{amsmath}

\newcommand{\Rp}{\ensuremath{R_p}\xspace}
\newcommand{\Rstar}{\ensuremath{R_{\star}}\xspace}

\newcommand{\Mstar}{\ensuremath{M_{\star}}}

\newcommand{\Mearth}{\ensuremath{M_{\oplus}}}
\newcommand{\Rearth}{\ensuremath{R_{\oplus}}}
\newcommand{\Rj}{\ensuremath{R_\mathrm{J}}}
\newcommand{\Mj}{\ensuremath{M_\mathrm{J}}}
\newcommand{\Teff}{\ensuremath{T_\mathrm{eff}}\xspace}
\newcommand{\Teq}{\ensuremath{T_\mathrm{eq}}\xspace}
\newcommand{\logg}{\ensuremath{\log g}\xspace}
\newcommand{\Rsun}{\ensuremath{R_{\odot}}\xspace}
\newcommand{\Msun}{\ensuremath{M_{\odot}}\xspace}
\newcommand{\vsini}{\ensuremath{v \sin i}\xspace}
\newcommand{\ms}{m s$^{-1}$\xspace}
\newcommand{\kms}{km s$^{-1}$\xspace}
\newcommand{\candtwo}{HD 89345\xspace}
\newcommand{\candthree}{HD 286123\xspace}
\newcommand{\pltwo}{HD 89345b\xspace}
\newcommand{\plthree}{HD 286123b\xspace}

\newcommand{\bjdtdb}{\ensuremath{\rm {BJD_{TDB}}}}
\newcommand{\feh}{\ensuremath{\left[{\rm Fe}/{\rm H}\right]}}

\newcommand{\msun}{\ensuremath{\,M_\Sun}}
\newcommand{\rsun}{\ensuremath{\,R_\Sun}}
\newcommand{\lsun}{\ensuremath{\,L_\Sun}}
\newcommand{\mj}{\ensuremath{\,M_{\rm J}}}
\newcommand{\rj}{\ensuremath{\,R_{\rm J}}}

\newcommand{\fave}{\langle F \rangle}
\newcommand{\fluxcgs}{10$^9$ erg s$^{-1}$ cm$^{-2}$}

\newcommand\sagan{$\ddagger$}
\newcommand\texaco{$\star$}
\newcommand\nsf{$\dagger$}
\newcommand\hub{\S}

\shorttitle{Two bright \ktwo planets}
\shortauthors{Yu et al.}

\begin{document}


\title{Two warm, low-density sub-Jovian planets orbiting bright stars in \\ \ktwo campaigns 13 and 14 }


\author{
Liang Yu\altaffilmark{1},
Joseph E.\ Rodriguez\altaffilmark{2},
Jason D.\ Eastman\altaffilmark{2},
Ian J.\ M.\ Crossfield\altaffilmark{1},
Avi Shporer\altaffilmark{1},
B.\ Scott Gaudi\altaffilmark{3},
Jennifer Burt\altaffilmark{1},
Benjamin J.\ Fulton\altaffilmark{4,\texaco},
Evan Sinukoff\altaffilmark{5,6}, 
Andrew W.\ Howard\altaffilmark{5},
Howard Isaacson\altaffilmark{7},
Molly R.\ Kosiarek\altaffilmark{8,\nsf},
David R.\ Ciardi\altaffilmark{9},
Joshua E.\ Schlieder\altaffilmark{10},
Kaloyan Penev\altaffilmark{11},
Andrew Vanderburg\altaffilmark{12,\sagan},
Keivan G.\ Stassun\altaffilmark{13},
Allyson Bieryla\altaffilmark{2},
R.\ Paul Butler\altaffilmark{14},
Perry Berlind\altaffilmark{2},
Michael L.\ Calkins\altaffilmark{2}, 
Gilbert A.\ Esquerdo\altaffilmark{2},
David W.\ Latham\altaffilmark{2},
Gabriel Murawski\altaffilmark{15},
Daniel J.\ Stevens\altaffilmark{3},
Erik A.\ Petigura\altaffilmark{5,\hub}, 
Laura Kreidberg\altaffilmark{2},
Makennah Bristow\altaffilmark{16}
}

\altaffiltext{1}{Department of Physics, and Kavli Institute for Astrophysics and Space Research, Massachusetts Institute of Technology, Cambridge, MA 02139, USA}
\altaffiltext{2}{Harvard-Smithsonian Center for Astrophysics, Cambridge, MA 02138, USA}
 \altaffiltext{3}{Department of Astronomy, The Ohio State University, Columbus, OH 43210, USA}
 \altaffiltext{4}{Department of Geology and Planetary Sciences, California Institute of Technology, Pasadena, CA 91125, USA}
 \altaffiltext{5}{Cahill Center for Astrophysics, California Institute of Technology, Pasadena, CA 91125, USA}
 \altaffiltext{6}{Institute for Astronomy, University of Hawai`i at M\={a}noa, Honolulu, HI 96822, USA}
 \altaffiltext{7}{Department of Astronomy, University of California, Berkeley, CA 94720, USA}
 \altaffiltext{8}{Department of Astronomy and Astrophysics, University of California, Santa Cruz, CA 95064, USA}
\altaffiltext{9}{NASA Exoplanet Science Institute, California Institute of Technology, Pasadena, CA 91125, USA}
\altaffiltext{10}{NASA Goddard Space Flight Center, 8800 Greenbelt Road, Greenbelt, MD 20771, USA}
\altaffiltext{11}{Department of Physics, The University of Texas at Dallas, 800 West Campbell Road, Richardson, TX 75080, USA}
\altaffiltext{12}{Department of Astronomy, The University of Texas at Austin, Austin, TX 78712, USA}
\altaffiltext{13}{Vanderbilt University, Department of Physics \& Astronomy, 6301 Stevenson Center Ln., Nashville, TN 37235, USA}
\altaffiltext{14}{Department of Terrestrial Magnetism, The Carnegie Institution for Science, NW Washington, DC 20015, USA}
\altaffiltext{15}{Gabriel Murawski Private Observatory (SOTES), Suwa\l ki, Poland}
\altaffiltext{16}{Department of Physics, University of North Carolina at Asheville, Asheville, NC 28804, USA}
\altaffiltext{\texaco}{Texaco Fellow}
\altaffiltext{\nsf}{NSF Graduate Research Fellow}
\altaffiltext{\sagan}{NASA Sagan Fellow}
\altaffiltext{\hub}{NASA Hubble Fellow}

\begin{abstract}
We report the discovery of two planets transiting the bright stars \candtwo (EPIC 248777106, $V=9.376$, $K=7.721$) in \ktwo Campaign 14 and \candthree (EPIC 247098361, $V=9.822$, $K=8.434$) in \ktwo Campaign 13. Both stars are G-type stars, one of which is at or near the end of its main sequence lifetime, and the other that is just over halfway through its main sequence lifetime. \candtwo hosts a warm sub-Saturn (0.66 \Rj, 0.11 \Mj, $\Teq=1100$ K) in an 11.81-day orbit. The planet is similar in size to WASP-107b, which falls in the transition region between ice giants and gas giants. \candthree hosts a Jupiter-sized, low-mass planet (1.06 \Rj, 0.39 \Mj, $\Teq=1000$ K) in an 11.17-day, mildly eccentric orbit, with $e=0.255\pm0.035$. Given that they orbit relatively evolved main-sequence stars and have orbital periods longer than 10 days, these planets are interesting candidates for studies of gas planet evolution, migration, and (potentially) re-inflation.  Both planets have spent their entire lifetimes near the proposed stellar irradiation threshold at which giant planets become inflated, and neither shows any sign of radius inflation. They probe the regime where inflation begins to become noticeable and are valuable in constraining planet inflation models. In addition, the brightness of the host stars, combined with large atmospheric scale heights of the planets, makes these two systems favorable targets for transit spectroscopy to study their atmospheres and perhaps provide insight into the physical mechanisms that lead to inflated hot Jupiters. 
\end{abstract}


\bibliographystyle{apj_ads}

\section{Introduction}

Giant planets have historically been an important class of transiting exoplanets, and many questions have been raised about their formation and evolution. The discovery of the first hot Jupiters immediately upended all existing giant planet formation models, which were based on observations of the Solar System. One of the most pressing open questions is how hot Jupiters, or Jupiter-mass planets orbiting at only a few percent of an astronomical unit from their host stars, are able to reach such short orbital periods. Although {\it in situ} formation has been considered as a possibility \citep[e.g.][]{bodenheimer00, batygin16}, hot Jupiters are most commonly thought to have formed at large radial distances and subsequently migrated inward to their present orbits. There have been several theories attempting to explain hot Jupiter migration. Some invoke interactions with a planetary or stellar companion: the gas giant planet is first injected into an eccentric orbit, which then undergoes tidal circularization \citep[e.g.][]{rasioford, fabryckytremaine}. Other theories suggest processes where the gas giant planet gradually moves inward by interacting with the protoplanetary disk, during which the orbit is kept circular \citep[e.g.][]{lin96, alibert05}. The two theories predict different orbital eccentricities and stellar obliquities as the planet migrates inward, yet it appears that stellar obliquities in hot Jupiter systems may be erased by tides raised by the planet on the star \citep[e.g.][]{schlaufman10, winn10}. We would then expect warm Jupiters - gas giants with orbital periods of 10 days or longer - which experience weaker tidal effects, to have retained the obliquity they had when emplaced in their current orbits. In reality, however, the interpretation is not that simple, as \citet{mazeh15} found that warm Jupiters seem to be showing effects of tidal realignment even at orbital distances where tidal effects should be negligible.

Another long-standing mystery is the anomalously large radii of ``inflated'' close-in giant planets. Many of the known transiting hot Jupiters have radii larger than expected by standard models of giant planets \citep[see, e.g., ][]{burrows97, bodenheimer01,guillotshowman}. Dozens of inflated hot Jupiters with radii $>1.2\ \Rj$ have been observed to orbit stars several Gyr old \citep{guillotgautier}. Although very young planets ($<$ 10 Myr) are expected to have radii this large, it is unclear how such inflated planets can exist around mature main sequence and even evolved stars \citep[e.g.][]{grunblatt17}.

Various mechanisms have been proposed to explain the large radii of hot Jupiters. Following \citet{lopezfortney}, the suggested mechanisms for inflating gas giants can be divided into two categories: in class I mechanisms, stellar irradiation incident on a planet is transported into the planet's deep interior, driving adiabatic heating of the planet and causing it to expand \citep[e.g.][]{batyginstevenson, arrassocrates}; in class II mechanisms, the inflationary mechanism simply acts to slow radiative cooling through the atmosphere, allowing a planet to retain its initial heat and inflated radius from formation \citep[delayed contraction, e.g.][]{burrows07}. The observation that the radii of giant planets increase with incident stellar irradiation hints that giant planet inflation is intimately linked to irradiation \citep{burrows00, bodenheimer01, lopezfortney}. We can distinguish between these two classes of models by studying warm Jupiters around stars that have recently evolved off the main sequence \citep[e.g.,][]{shporer17, smith17}.  The irradiation levels experienced by warm Jupiters around main sequence stars are not high enough to cause inflation, but as their host stars move up the subgiant and red-giant branches, they will experience enormous increases in their irradiation levels. If class I mechanisms are responsible for giant planet inflation, then warm Jupiters should inflate in response to the increased irradiation \citep{assef2009,Spiegel:2012,hartman16}. On the other hand, an exclusively non-inflated population of warm Jupiters around evolved stars would favor class II mechanisms \citep{lopezfortney}.

Finally, we have yet to even understand the formation mechanism of giant planets. The positive correlation between the fraction of stars with short-period giant planets and stellar metallicity hints that planets form through core accretion \citep[e.g.][]{santos04, johnson10}. In the core accretion scenario, a rocky core forms through the coagulation of planetesimals; when the mass of the gaseous envelope relative to the solid core mass reaches a critical ratio, rapid gas accretion occurs and a giant planet is formed \citep[e.g.][]{pollack96}. Gas accretion is expected to start in the mass regime between Neptune and Saturn \citep{mordasini15}, the transition zone between ice giants and gas giants. Yet this regime is not very well understood given the small number of known planets that fall within this mass range. In particular, the core accretion model struggles to explain why ice giants do not undergo the runaway gas accretion that would have turned them into gas giants \citep{helledbodenheimer}.

In this paper, we present the discovery of two exoplanets observed by \ktwo, which are pertinent to the problems described above: one sub-Saturn transiting a bright star \candtwo (EPIC 248777106), and a warm Saturn orbiting a similarly bright star \candthree (EPIC 247098361), with both stars well into or nearing the ends of their main sequence lifetimes. Despite their large radii, both planets have low masses, which make them promising targets for atmospheric characterization. They are also interesting additions to the currently available set of giant planets to study radius inflation, which consists primarily of Jupiter-massed objects. We describe our discovery and observations in Section~\ref{sec:obs}, our derivation of stellar and planetary parameters in Sections~\ref{sec:stars} and~\ref{sec:planet}, and potential implications for giant planet migration, inflation and formation theories in Section~\ref{sec:conclusions}.

\section{Observations}
\label{sec:obs}
\candthree was proposed as a \ktwo target in Campaign 13 (C13) in four programs: GO13071 (PI Charbonneau), GO13122 (PI Howard), GO13024 (PI Cochran) and GO13903 (GO Office).  \candtwo was proposed as a target in Campaign 14 (C14) in five programs: GO14010 (PI Lund), GO14009 (PI Charbonneau), GO14028 (PI Cochran), GO14021 (PI Howard) and GO14901 (GO Office). C13 was observed from 2017 Mar 08 to May 27, and C14 was observed from 2017 Jun 01 to Aug 19. \candtwo and \candthree's photometric and spectroscopic properties are given in Table~\ref{tab:stellar}.

\begin{deluxetable*}{l l l l}[bt]
\hspace{-1in}\tabletypesize{\scriptsize}
\tablecaption{  Stellar parameters of \candtwo and \candthree from the literature and spectroscopy}
\tablewidth{0pt}
\tablehead{
\colhead{Parameter}  & \colhead{\candtwo} & \colhead{\candthree} & \colhead{Source}
}
\startdata
\multicolumn{4}{l}{\hspace{1cm}\em Identifying Information} \\
$\alpha_{\rm J2000}$ R.A. (hh:mm:ss)  & 10:18:41.06 & 04:55:03.96 & \\
$\delta_{\rm J2000}$ Dec. (dd:mm:ss) & 10:07:44.5 & 18:39:16.33 & \\
Other identifiers
  & TYC 840-840-1 & TYC 1284-745-1 \\
  & 2MASS J10184106+1007445 & 2MASS J04550395+1839164 \\
  & EPIC 248777106 & EPIC 247098361 & \\
  & K2-234 & K2-232 \\
 \ktwo campaign & 14 & 13 & \\
\multicolumn{4}{l}{\hspace{1cm}\em Photometric Properties} \\
B (mag)..........  &  10.148 $\pm$ 0.039 & 10.520 $\pm$ 0.051 & 1 \\
V (mag) .......... &  9.376 $\pm$ 0.028 & 9.822 $\pm$ 0.038 & 1 \\
J (mag)..........  & 8.091 $\pm$ 0.020 & 8.739 $\pm$ 0.030 & 2\\
H (mag) .........  & 7.766 $\pm$ 0.040 & 8.480 $\pm$ 0.018 & 2 \\
Ks (mag) ........  & 7.721 $\pm$ 0.018 & 8.434 $\pm$ 0.017 & 2 \\
W1 (mag) ........  & 7.763 $\pm$ 0.028 & 8.380 $\pm$ 0.024 & 3\\
W2 (mag) ........  & 7.759 $\pm$ 0.020 & 8.419 $\pm$ 0.019 & 3\\
W3 (mag) ........  & 7.729 $\pm$ 0.019 & 8.391 $\pm$ 0.027 & 3 \\
\multicolumn{4}{l}{\hspace{1cm}\em Spectroscopic and Derived Properties} \\
Spectral Type  & G5V-G6V  & F9V-G0V & 4\\
$\mu_{\alpha}$ (mas~yr$^{-1}$) & 5.348 $\pm$ 0.079 & 62.064 $\pm$ 0.077 & 5 \\
$\mu_{\delta}$ (mas~yr$^{-1}$) & -42.449 $\pm$ 0.071 & -48.245 $\pm$ 0.051 & 5 \\
Parallax (mas) &  7.528 $\pm$ 0.046 & 7.621 $\pm$ 0.044 & 5 \\
Barycentric RV (km~s$^{-1}$) &  2.4 $\pm$ 0.1 & 22.4 $\pm$ 0.1 & TRES; this paper \\
\vsini\ (km~s$^{-1}$) & 3 $\pm$ 1 & 3 $\pm$ 1 & APF; this paper \\
Space motion ($U^{*},V,W$) (km~s$^{-1}$) & (21.5$\pm$0.1, -9.8$\pm$0.1, 1.5$\pm$0.1) & (-14.9$\pm$0.1, -34.5$\pm$0.3, 13.5$\pm$0.1)  & this paper
\enddata
\label{tab:stellar}
\tablenotetext{}{{\bf References:} (1) \citet{tycho}; (2) \citet{twomass}; (3) \citet{allwise}; (4) \citet{pecautmamajek13}, (5) \citet{gaia16, gaia:2018}. $^{*}$Positive $U$ is in the direction of the Galactic center }
\end{deluxetable*}

\subsection{\ktwo Photometry}
We converted the processed \ktwo target pixel files into light curves using an approach identical to that described in \citet{crossfield15}. In brief, we computed the raw photometry by summing the flux within a soft-edged circular aperture centered around the target star, and used the publicly available \texttt{k2phot} photometry code\footnote{https://github.com/petigura/k2phot} to model out the time- and roll-dependent variations with a Gaussian process. We then used the publicly available \texttt{TERRA} algorithm\footnote{https://github.com/petigura/terra} \citep{petigura13a, petigura13b} to search for transit-like events and manually examined diagnostic plots for all signals with S/N $\geq 10$. \texttt{TERRA} identified a planet candidate orbiting \candtwo with $P=11.81$ days and S/N $=24$ in Campaign 14, and another candidate orbiting \candthree with $P=11.17$ days and S/N $=495$ in Campaign 13. 

\begin{figure*}[htb]

\centerline{%
\includegraphics[width=0.5\textwidth]{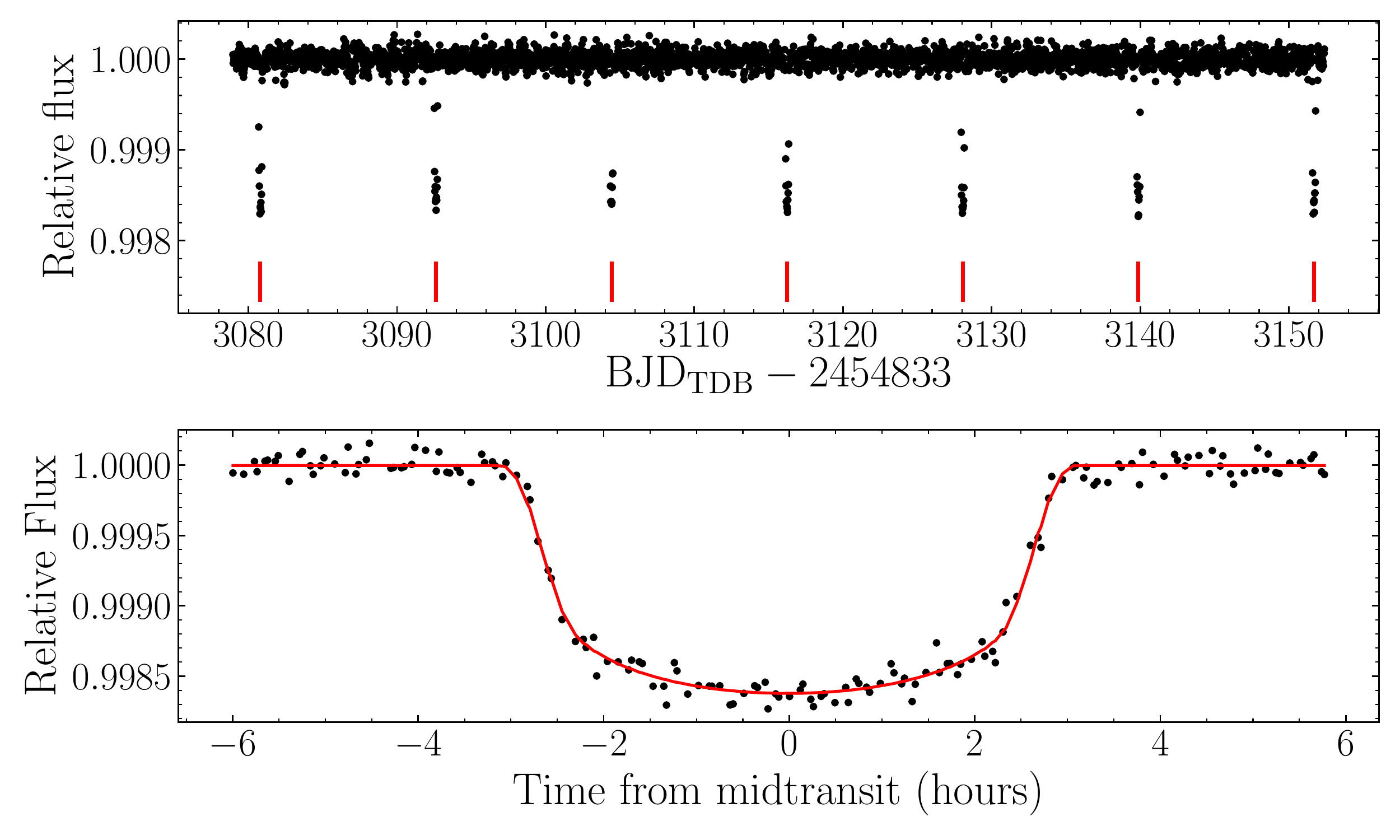}%
\includegraphics[width=0.5\textwidth]{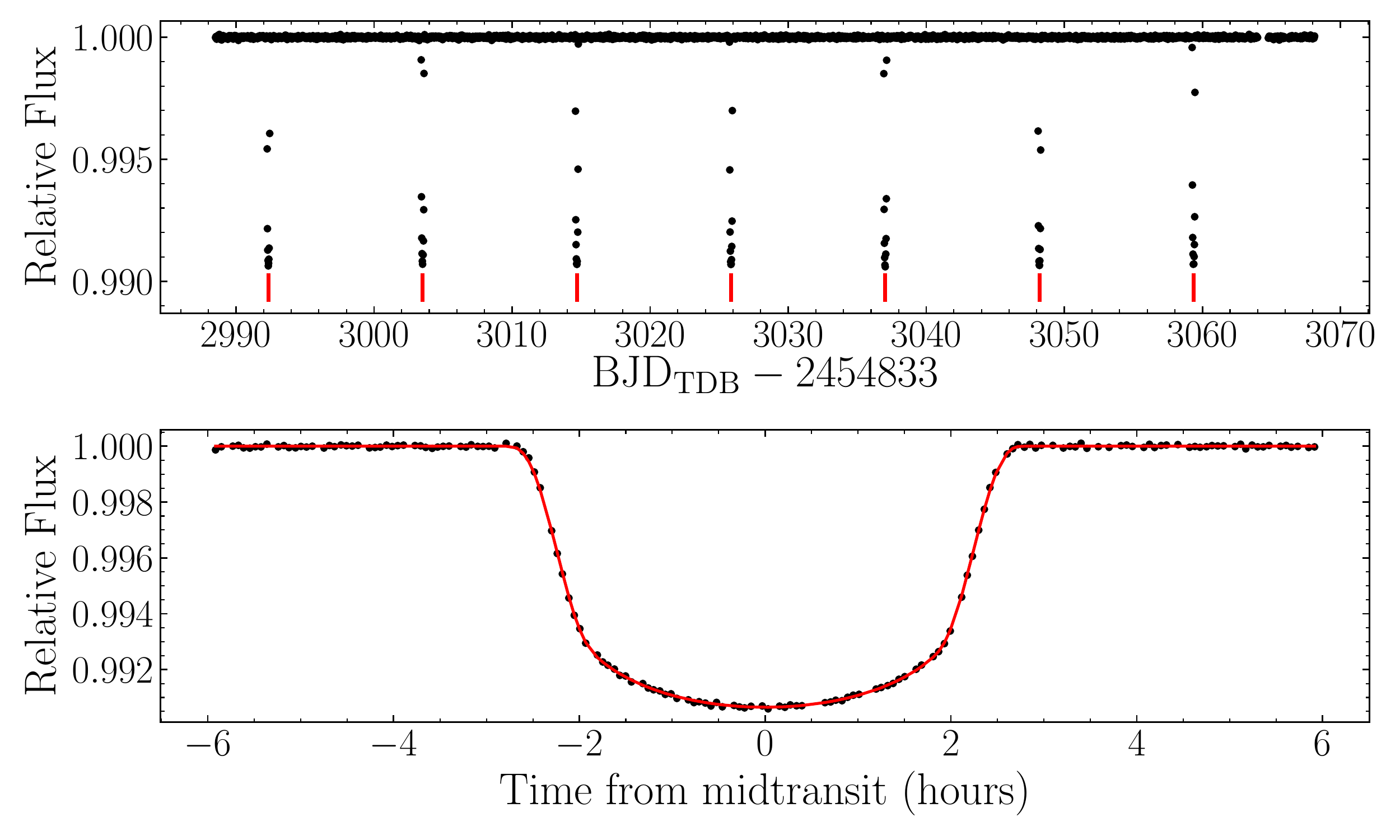}%
}%
\caption{{\it Left}: Calibrated \ktwo photometry for \candtwo (top), with vertical ticks indicating the locations of the transits, and phase-folded photometry and best-fit light curve model (bottom). {\it Right:} Same, but for \candthree.}
\label{fig:lcs} 
\end{figure*}

After identifying the transits, we produced new light curves by simultaneously fitting the transits, the \ktwo roll systematics, and long-timescale stellar/instrumental variability. Reprocessing the \ktwo light curves in this way prevents the shape of the transits from being biased by the removal of \ktwo systematics. We used light curves and systematics corrections derived using the method of \citet{vj14} as initial guesses for our simultaneous fits, which we then performed following \citet{v16}. Throughout the rest of this paper, we use these simultaneously-fit light curves in our analysis and our plots. Fig.~\ref{fig:lcs} shows the flattened\footnote{We flattened the light curves by dividing away the best-fit long-timescale variability from our simultaneously-fit light curve.} and detrended light curves of \candtwo and \candthree. 

\subsection{Ground-based Followup}
In this section, we present our ground-based photometric and spectroscopic observations used to confirm the planetary nature of \pltwo and \plthree.

\subsubsection{Spectroscopic followup}
We used the HIRES spectrograph \citep{hires} at the W. M. Keck Observatory to measure high-resolution optical spectra of the two targets. Observations and data reduction followed the standard procedures of the California Planet Search \citep[CPS;][]{Howard10}.  For both stars, the $0.86\arcsec{} \times$ 14\arcsec{} ``C2'' decker was placed in front of the slit and the exposures were terminated once an exposure meter reached 10,000 counts, yielding a signal-to-noise ratio (SNR) of 45 per pixel at 550 nm. Additional spectra of \candtwo were collected to measure precise radial velocities (RVs), by placing a cell of gaseous iodine in the converging beam of the telescope, just ahead of the spectrometer slit. The iodine cell is sealed and maintained at a constant temperature of 50.0 $\pm$0.1$^{\circ}$C to ensure that the iodine gas column density remains constant over decades. The iodine superimposes a rich forest of absorption lines on the stellar spectrum over the 500-620 nm region, thereby providing a wavelength calibration and proxy for the point spread function (PSF) of the spectrometer. Once extracted, each spectrum of the iodine region is divided into $\sim$700 chunks, each of which is 2 \AA\ wide. Each chunk produces an independent measure of the wavelength, point spread function, and Doppler shift, determined using the spectral synthesis technique described by \cite{Butler96}. The final reported Doppler velocity for a stellar spectrum is the weighted mean of the velocities of all the individual chunks.  The final uncertainty of each velocity is the weighted average of all 700 chunk velocities. These iodine exposures were terminated after 50,000 counts (SNR = 100 per pixel), typically lasting 2 min.  For both stars, a single iodine-free ``template'' spectrum with a higher SNR of 225 was also collected using the narrower ``B3''  decker ($0.57\arcsec{} \times$ 14\arcsec{}). RVs were measured using the standard CPS Doppler pipeline \citep{Marcy92, Valenti95, Butler96, Howard09}. Each observed spectrum was forward modeled as the product of an RV-shifted iodine-free spectrum and a high-resolution/high-SNR iodine transmission spectrum convolved with a PSF model.  Typical internal RV uncertainties were 1.5 m~s$^{-1}$.

We obtained additional spectra for the two targets with the Tillinghast Reflector Echelle Spectrograph \citep[TRES;][]{tres} on the 1.5 m telescope at the Fred L. Whipple Observatory on Mt. Hopkins, AZ. The TRES spectra have a resolution of ~44,000 and were extracted as described in \citet{buchhave2010}. We obtained 8 TRES spectra of \candthree in Oct 2017. The average S/N per resolution element (SNRe) was 46, which was determined at the peak continuum of the Mgb region of the spectrum near 519 nm. \candtwo was observed twice, once in 2017 Nov and again in 2017 Dec, with an average SNR of 54. The TRES spectra were not used to determine an orbital solution but were used to determine stellar parameters (see Section~\ref{sec:stars}).

The RV dataset for \candthree is comprised of 19 velocities obtained between 2018 Oct and 2019 Feb using the Automated Planet Finder (APF), a 2.4m telescope located atop Mt. Hamilton at Lick Observatory. The telescope is paired with the Levy echelle spectrograph, and is capable of reaching 1 m s$^{-1}$ precision on bright, quiet stars. The Levy spectrograph is operated at a resolution of $\sim$90,000 for RV observations and covers a wavelength range of 370-900 nm, though only the 500-620 nm Iodine region is used in extracting Doppler velocities \citep{vogt14}. APF RVs were collected using the same iodine-based methodology described above.

The APF is a dedicated exoplanet facility, and employs a dynamic scheduler to operate without the aid of human observers \citep{burt15}. Due to the large expected RV semi-amplitude of the transiting planet (K $\sim$26 m s$^{-1}$) and the desire to use the telescope as efficiently as possible, we set the desired RV precision in the dynamic scheduler to 4 m s$^{-1}$. The exposure times necessary to achieve this precision were automatically calculated in real time to account for changing atmospheric conditions, and the resulting RV data set has a mean internal uncertainty of 3.9 m s$^{-1}$. In our analysis, we chose to omit data taken on one night due to a low number of photons in the iodine region, caused by cloudy observing conditions. 

APF RVs for \candtwo were collected in exactly the same way between Nov 2017 and Feb 2018 except that we used fixed 30-min exposure times giving SNR $\sim 80$. We also utilized the iodine-free ``template" observations collected on Keck during the RV extraction in order to avoid duplication of data. All RV measurements for the two systems are reported in Tables~\ref{tab:rv1} and~\ref{tab:rv2}.

\begin{deluxetable}{l l l l}[bt]
\tabletypesize{\scriptsize}
\tablecaption{  Radial Velocities for \candtwo \label{tab:rv1}}
\tablewidth{0pt}
\tablehead{
\colhead{\bjdtdb} & \colhead{RV (m s$^{-1}$)} & \colhead{$\sigma_{RV}$ (m s$^{-1}$)} & \colhead{Instrument}
}
\startdata
2458088.069276 & 9.1 & 2.5 & APF \\
2458089.027427 & 14.0 & 2.4 &  APF \\
2458092.004495 & -7.8 & 1.9 & HIRES\\
2458093.080794 & -4.4 & 4.8 & APF \\
2458094.982520 & -2.0 & 2.5 & APF \\
2458099.986259 & 5.6 & 1.7 & HIRES\\
2458107.007150 & -9.6 & 2.4 & APF \\
2458109.930945 & -6.0 & 3.5 & APF \\
2458113.049078 & 2.9 & 1.4 & HIRES\\
2458114.005711 & 6.4 & 2.8 & APF \\
2458114.048573 & 2.3 & 1.7 & HIRES\\
2458115.846145 & -0.5 & 3.6 & APF \\
2458116.934534 & -8.5 & 1.2 & HIRES\\
2458118.934942 & -8.0 & 1.6 & HIRES\\
2458120.050818 & -8.9 & 2.8 & APF \\
2458125.024846 & 1.0 & 1.7 & HIRES \\
2458161.055191 & 1.7 & 1.2 & HIRES \\
2458181.912190 & 5.3 & 1.6 & HIRES \\
2458194.946988 & 14.0 & 1.6 & HIRES  \\
2458199.783821 & -14.7 & 1.8 & HIRES \\
2458209.952119 & -0.3 & 1.8 & HIRES
\enddata 
\end{deluxetable}

\begin{deluxetable}{l l l l}[bt]
\tabletypesize{\scriptsize}
\tablecaption{  Radial Velocities for \candthree \label{tab:rv2}}
\tablewidth{0pt}
\tablehead{
\colhead{\bjdtdb} & \colhead{RV (m s$^{-1}$)} & \colhead{$\sigma_{RV}$ (m s$^{-1}$)} & \colhead{Instrument}
}
\startdata
2458054.788304514 & -0.6 & 3.1 &  APF\\
2458055.777965865 & 8.7 & 3.1 &  APF\\
2458070.727222747 & -28.0 & 3.6 &  APF\\
2458076.725029452 & -2.1 & 4.7 &  APF\\
2458079.684549427 & 1.4 & 2.9 &  APF\\
2458085.797267544 & -30.0 & 3.0 &  APF\\
2458089.686288970 & 2.6 & 2.8 &  APF\\
2458097.641321792 & -14.0 & 4.2 &  APF\\
2458098.690303951 & -7.8 & 3.2 &  APF\\
2458099.706504478 & 1.1 & 2.7 &  APF\\
2458100.616924353 & 29.3 & 8.7 &  APF\\
2458102.620537149 & 4.8 & 3.2 &  APF\\
2458114.740644800 & -13.0 & 4.0 &  APF
\enddata 
\end{deluxetable}

\subsubsection{Keck/NIRC2 Adaptive Optics Imaging}
\label{sec:ao}
\begin{figure*}[htb]
\centerline{%
\includegraphics[width=0.49\textwidth, trim={2.5cm 2cm 2.5cm 2cm},clip]{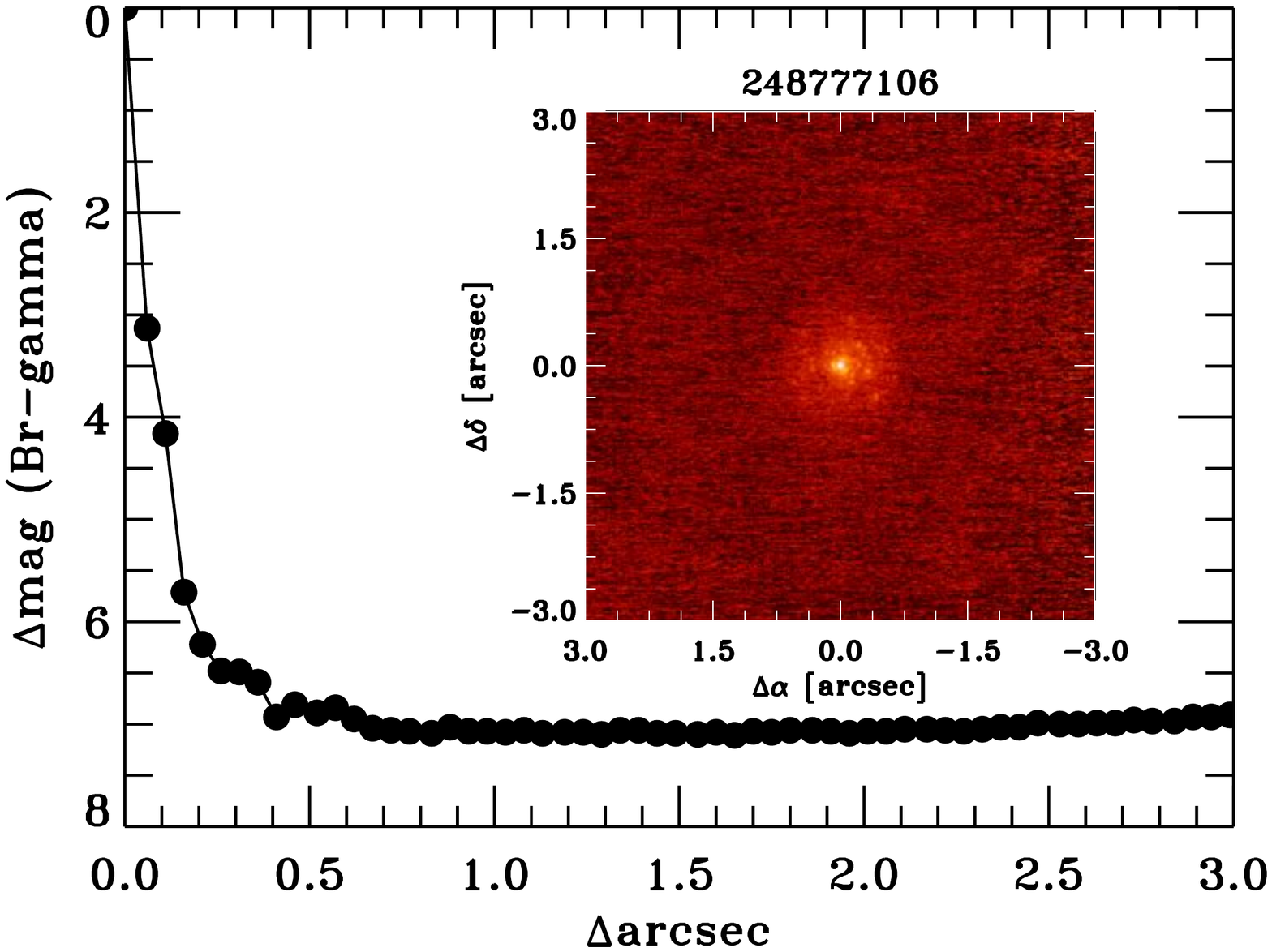}%
\includegraphics[width=0.49\textwidth, trim={2.5cm 2cm 2.5cm 2cm},clip]{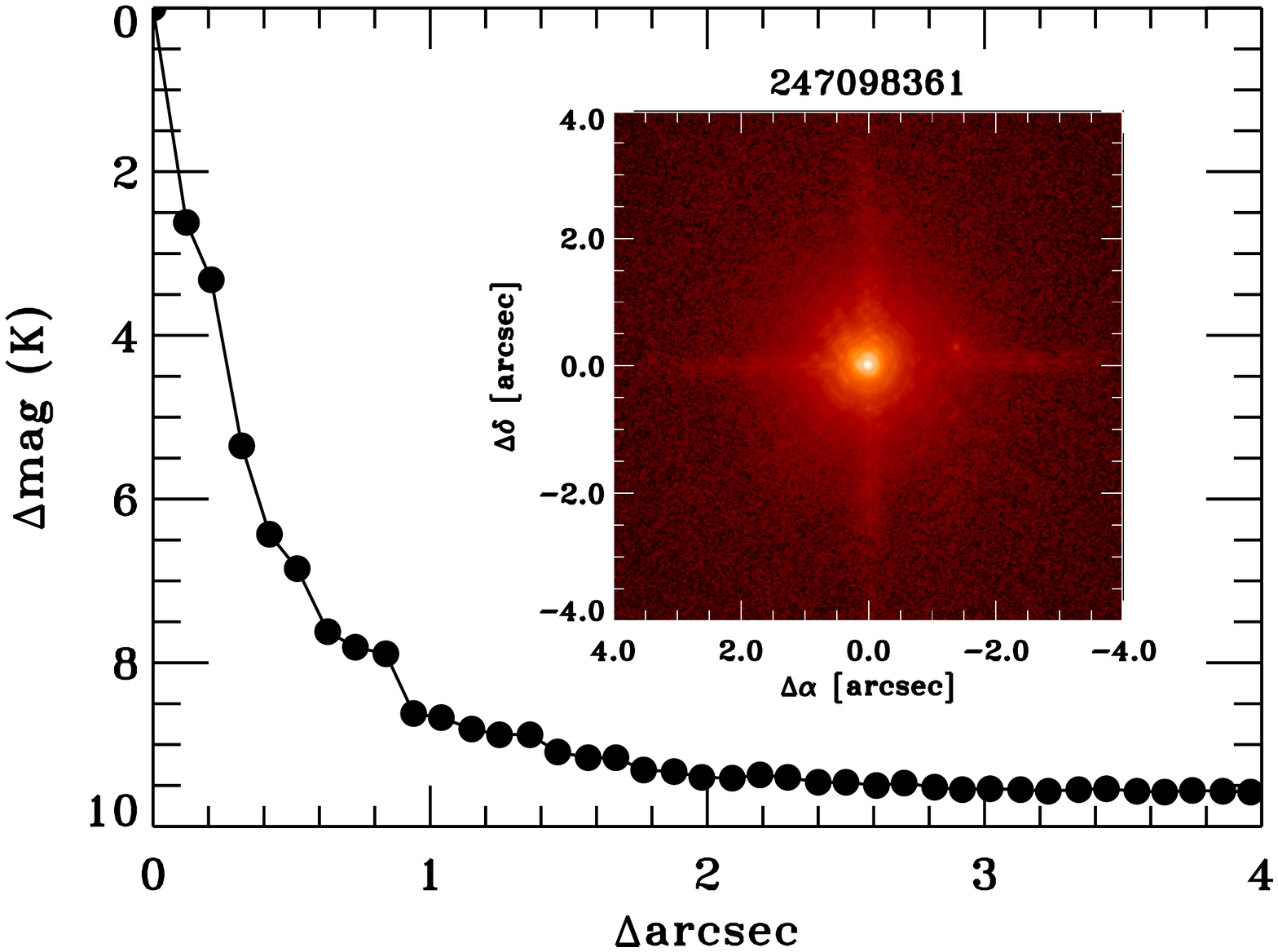}%
}%
\caption{AO images (inset) and $K_S$-band contrast curves for \candtwo (left) and \candthree (right). \candtwo was imaged with Keck/NIRC2, and \candthree was imaged with Palomar/PHARO. The right image shows a faint companion at about $\sim 1.4\arcsec$ away from \candthree, but this cannot be the source of the observed transit signals (see Section~\ref{sec:ao}).}
\label{fig:ao} 
\end{figure*}

We obtained NIR adaptive optics (AO) imaging of \candtwo through clear skies with $\sim0.8\arcsec$ seeing on the night of 2017 Dec 29 using the 10m Keck II telescope at the W. M. Keck Observatory. The star was observed behind the natural guide star AO system using the NIRC2 camera in narrow angle mode with the large hexagonal pupil. We observed using the narrow-band Br-$\gamma$ filter ($\lambda_c$ = 2.1686 $\mu$m; $\Delta\lambda$ = 0.0326 $\mu$m) with a 3-point dither pattern that avoids the noisier lower left quadrant of the NIRC2 detector. Each dither was offset from the previous position by 0.5$^{\prime\prime}$ and the star was imaged at 9 different locations across the detector. The integration time per dither was 1s for a total time of 9s. The narrow angle mode of NIRC2 provides a field-of-view of $10\arcsec$ and a plate scale of about 0.01$^{\prime\prime}$ pixel$^{-1}$. We used the dithered images to remove sky-background, then aligned, flat-fielded, dark subtracted and combined the individual frames into a final combined image (see Fig.~\ref{fig:ao} inset). The final images had a FWHM resolution of $\sim$60 mas, near the diffraction limit at $\sim$2.2 $\mu$m.

 We also obtained NIR high-resolution AO imaging of \candthree at Palomar Observatory with the $200\arcsec$ Hale Telescope at Palomar Observatory on 2017 Sep 06 using the NIR AO system P3K and the infrared camera PHARO \citep{hayward01}. PHARO has a pixel scale of $0.025\arcsec$ per pixel with a full field of view of approximately $25\arcsec$. The data were obtained with a narrow-band $Br$-$\gamma$ filter $(\lambda_o = 2.166; \Delta\lambda = 0.02 \ \mu$m ). The AO data were obtained in a 5-point quincunx dither pattern with each dither position separated by 4$^{\prime\prime}$.  Each dither position is observed 3 times with each pattern offset from the previous pattern by $0.5^{\prime\prime}$ for a total of 15 frames.  The integration time per frame was 9.9s for a total on-source time of 148.5s. We use the dithered images to remove sky background and dark current, and then align, flat-field, and stack the individual images. The PHARO AO data have a resolution of 0.10$^{\prime\prime}$ (FWHM). 

To determine the sensitivity of the final combined images, we injected simulated sources at positions that were integer multiples of the central source FWHM scaled to brightnesses where they could be detected at 5$\sigma$ significance with standard aperture photometry.  We compared the $\Delta$-magnitudes of the injected 5$\sigma$ sources as a function of their separation from the central star to generate contrast sensitivity curves (Fig.~\ref{fig:ao}). We were sensitive to close companions and background objects with $\Delta$Br-$\gamma$ $\approx$ 6 at separations $\ge$200 mas. No additional sources were detected down to this limit in the field-of-view of \candtwo, and the target appears single at the limiting resolution of the images. 

A stellar companion was detected near \candthree in the $Br$-$\gamma$ filter with PHARO. The companion separation was measured to be $\Delta\alpha = -1.39\arcsec \pm 0.01\arcsec$ and $\Delta\delta = 0.28\arcsec \pm 0.03\arcsec$.  The companion has a measured differential brightness in comparison to the primary star of $\Delta_K = 6.75\pm0.05$ mag, which implies deblended stellar 2MASS $K$-band magnitudes of $K_S = 8.45\pm0.02$ mag and $K_S = 15.2\pm0.1$ mag for the primary and the companion respectively.  Utilizing Kepler magnitude ($K_p$)-$K_S$ relationships from \citet{howell12}, we derive approximate deblended Kepler magnitudes of $K_p = 9.81\pm0.01$ mag for the primary and $K_p = 17.3\pm0.8$ mag for the companion. The resulting Kepler magnitude difference is $\Delta K_p = 7.5\pm0.8$ mag. The companion star therefore cannot be responsible for the transit signals, but is potentially a bound stellar companion.  At a separation of 1.4\arcsec, the projected separation of the companion is approximately 175 AU. This translates to an orbital period of about a $\sim$2300 yr ($10^6$ days), which is near the peak of the period distribution of binaries \citep{raghavan2010} and within the $80\%$ likelihood of AO-detected companions being bound for these separations \citep{hirsch2017}.  With an infrared magnitude difference of $\Delta_K = 6.75$ mag and assuming the distance of \candthree, the companion star has an infrared magnitude similar to that of an M7V dwarf \citep{pecautmamajek13}.  

The AO imaging rules out the presence of any additional stars within $\sim 0.5$\arcsec\ of \candthree ($\sim 30$~AU) and the presence of any brown dwarfs, or widely-separated tertiary components down to $K_S=16.4$ beyond 0.5\arcsec ($\sim 30-1000$~AU). All data and sensitivity curves are available on the ExoFOP-K2 site\footnote{https://exofop.ipac.caltech.edu}.

We also searched for any faint sources within the \ktwo apertures used but beyond the field of view of the AO imaging by examining archival images from imaging surveys including SDSS9, 2MASS, Pan-STARRS and DECaLS/DR3, and catalogs including UCAC, GSC2.3, 2MASS and SDSS12. Across all surveys and catalogs, we identified no sources brighter than 19 magnitudes in the $g'$-band and 18 magnitudes in the $r'$-band within 40$^{\prime\prime}$ of either star. The optical flux contribution of any faint companion is below the precision of \ktwo and can be safely ignored in our transit fits.

\subsubsection{Ground-Based Photometry}
We obtained additional ground-based photometric observations of \candthree on the night of 2017 Sep 29. One of us (G.M.) observed the second half of the transit from Suwa\l ki, Poland using a 78mm ASI178MM-cooled camera with a 1/1.8\arcsec CMOS IMX178 sensor and Canon FD 300mm f/2.8 lens. The images have pixel scales of 1.65\arcsec/pixel. No filter was used, and each measurement consists of 100 binned 3s exposures.

\begin{figure}[hbt]
\centering
\includegraphics[width=0.45\textwidth]{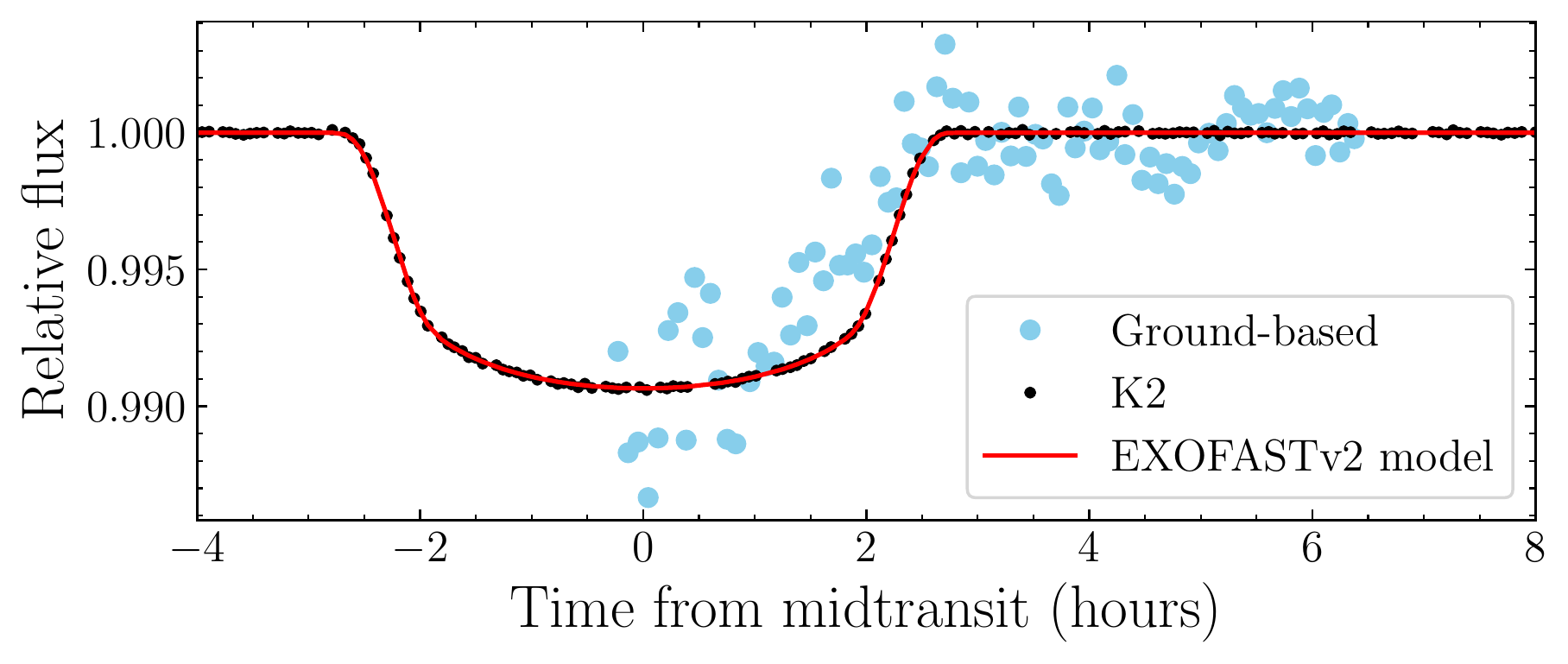}
\caption{Detrended ground-based light curve of \candthree (blue) and \ktwo light curve (black) phase-folded to the same ephemeris and overplotted with the transit model (red) from our global fit in Section~\ref{sec:exofast}.}
\label{fig:gblc} 
\end{figure}

Dark and flat calibrations were applied to each frame. The aperture used was a circular aperture with a radius corresponding to $8.7\arcsec$. Two stable stars within the field of view were used as reference stars, and the flux of \candthree was divided by the sum of the reference stars' fluxes. We modeled the out-of-transit variations with a quadratic function, which was also divided out to obtain the detrended light curve. Fig.~\ref{fig:gblc} shows the resulting light curve overplotted with the \ktwo light curve, phase-folded to the same ephemeris. The data clearly show the transit egress and so confirm the ephemeris
of this planet, but in the rest of our analysis we use only the \ktwo light curve.

\section{Host Star Characterization}
\label{sec:stars}
\subsection{Spectral Analysis} 
\label{spec}

We searched the iodine-free Keck/HIRES for spectroscopic blends using the algorithm of \citet{Kolbl15}, which is sensitive to secondary stars with $>1\%$ flux and $\Delta\mathrm{RV}$ $>$ 10 \kms relative to the primary star.  No secondary lines were detected in either spectrum.

We calculated initial estimates of the spectroscopic parameters of the host stars from our iodine-free Keck/HIRES spectra using the SpecMatch procedure \citep{petigura:2015phd}. SpecMatch searches a grid of synthetic model spectra \citep{coelho05} to fit for the effective temperature (\Teff), surface gravity ($\log g$), metallicity ([Fe/H]) and projected equatorial rotation velocity of the star ($v \sin i$). The resulting values are \Teff$=5532 \pm 100$ K, $\log g = 3.71 \pm 0.10$, [Fe/H]$=+0.44 \pm 0.06$, $v \sin i = 3 \pm 1$ km/s for \candtwo, and \Teff$=5909 \pm 100$ K, $\log g=4.25 \pm 0.10$, [Fe/H]$=+0.05 \pm 0.06$, $v \sin i =3 \pm 1$ km/s for \candthree. We adopt these values as starting points and/or priors for the \texttt{isoclassify} fits described in Section~\ref{iso} and the global fit described in Section~\ref{sec:exofast}.

As a consistency check, we also estimated the spectroscopic parameters using our TRES spectra and the Stellar Parameter Classification tool \citep[SPC;][]{buchhave12, buchhave14}. SPC works by cross-correlating observed spectra with a grid of synthetic model spectra generated from \citet{kurucz92} model atmospheres. From these fits, we obtained weighted averages of \Teff$=5676 \pm 50$ K, $\log g = 4.13 \pm 0.10$, [Fe/H]$=+0.50 \pm 0.08$, $v \sin i = 3.3 \pm 0.5$ km/s for \candtwo, and \Teff$=5877 \pm 53$ K, $\log g = 4.27 \pm 0.10$, [Fe/H]$=+0.03 \pm 0.08$, $v \sin i = 3.9 \pm 0.5$ km/s for \candthree. The values from SPC are in agreement with those from SpecMatch, except for the slightly higher $\log g$ value for \candtwo from SPC. Given that \candtwo is a slightly evolved star, spectroscopic \logg estimates are expected to be less reliable. As shown by \citet{torres12}, reliance on spectroscopically determined \logg can lead to considerable biases in the inferred evolutionary state, mass and radius of a star. Therefore we avoid imposing any priors on \logg for our global fit in Section~\ref{sec:exofast}.


\subsection{Evolutionary Analysis} 
\label{iso}
We then use the stellar parameters derived from HIRES spectra as well as broadband photometry and parallax as inputs for the grid-modeling method implemented in the stellar classification package \texttt{isoclassify} \citep{isoclassify}. \texttt{isoclassify} derives posterior distributions for stellar parameters (\Teff, $\log g$, [Fe/H], radius, mass, density, luminosity and age) through direct integration of isochrones from the MIST database \citep{dotter16,choi16} and synthetic photometry. Both target stars have parallaxes from Gaia DR2, but are saturated in the Sloan $z$ band. We therefore input for each star its 2MASS $JHK$ and Tycho $BV$ magnitudes, Gaia parallax, and \Teff, $\log g$, and [Fe/H] from SpecMatch. 
The $V$-band extinction $A_V$ is left as a free parameter. From this fit, we obtained $\Rstar=1.720\pm0.051 \Rsun$ and $\Mstar=1.147 \pm 0.034 \Msun$ for \candtwo and $\Rstar=1.214 \pm 0.043 \Rsun$ and $\Mstar = 1.063 \pm 0.047 \Msun$ for \candthree. These values are consistent with the final determined stellar parameters from our EXOFASTv2 global fit (See Table \ref{tbl:Parameters}). 


\subsection{UVW Space Motions, Galactic Coordinates, and Evolutionary States of the Host Stars}

To calculate the absolute radial velocities of the two host stars, we used the TRES observation with the highest SNR for each and corrected for the gravitational redshift by adding -0.61 km/s. This gives us an absolute velocity of $2.4~{\rm km~s^{-1}}$ for \candtwo and $22.4~{\rm km~s^{-1}}$ for \candthree. We quote an uncertainty of 0.1 km/s which is an estimate of the residual systematics in the IAU radial velocity standard star system. 

\subsubsection{\candtwo}

\candtwo is located at equatorial coordinates $\alpha=10^h18^m41\fs06$,
and $\delta=+10\arcdeg07\arcmin44\farcs5$ (J2000), which corresponds to Galactic coordinates of $\ell=230.8\arcdeg$ and $b=50.2\arcdeg$.  Given the Gaia distance of $\sim 127$~pc, \candtwo lies roughly 100 pc above the Galactic plane. Using the Gaia DR2 proper motion of $(\mu_\alpha,\mu_\delta)=(5.348 \pm 0.079, -42.449 \pm 0.071)~{\rm mas~yr}^{-1}$, the Gaia parallax, and the absolute radial velocity as determined from the TRES spectroscopy of $2.4 \pm 0.1~{\rm km~s^{-1}}$, we find
that \candtwo has a three-dimensional Galactic space motion of $(U,V,W)= (21.5 \pm 0.1, -9.8 \pm 0.1, 1.5 \pm 0.1)~{\rm km~s^{-1}}$, where positive $U$ is in the direction of the Galactic center, and we have adopted the \citet{Coskunoglu:2011} determination of the solar motion with respect to the local standard of rest. These values yield a 99.4\% probability that \candtwo is a thin disk star, according to the classification scheme of \citet{Bensby:2003}.

Note that stars of the mass of \candtwo ($M_*\sim 1.2~M_\odot$) that are close to the zero age main sequence typically have spectral types of roughly F5V-F8V \citep{pecautmamajek13}, but in fact \candtwo has a $T_{\rm eff}$ and colors that are more consistent with a much later spectral type of G5V-G6V \citep{pecautmamajek13}.  Furthermore, it has a radius of $R\sim 1.74~R_\odot$; much larger than one would expect of its mass if it were on the zero age main sequence.  All of this implies that \candtwo has exhausted or nearly exhausted its core hydrogen, and is currently in or close to the relatively short subgiant phase of its evolution, as it moves toward the giant branch.  The location of \candtwo above the disk \citep{Bovy:2017} and Galactic velocities are all consistent with this scenario. 

This conclusion is corroborated by the properties of the star inferred from the global fit to the transit, radial velocity, spectral energy distribution (SED), and parallax data described in Section~\ref{sec:exofast}.  A joint fit to these data measure, nearly directly and empirically, the stellar radius, density, surface gravity, and luminosity. As we note in Section~\ref{sec:exofast}, the global fit in fact yields two solutions, one on the main sequence and one on the subgiant branch. Together with the \Teff and \feh, we can locate both of these solutions on a ``theoretical" Hertzsprung-Russell diagram (see Figure~\ref{fig:yy}).  When comparing these values to MIST evolutionary tracks \citep{dotter16, choi16}, we infer that \candtwo has an age of either $\sim 4.2$~Gyr or $7.5$~Gyr and is indeed either near or just past the end of its main-sequence lifetime. 

\subsubsection{\candthree}

\candthree is located at equatorial coordinates $\alpha=4^h55^m03\fs9$,
and $\delta=+18\arcdeg39\arcmin16\farcs33$ (J2000), which correspond to the Galactic coordinates of $\ell=182.1\arcdeg$ and $b=-15.3\arcdeg$.  Given the Gaia distance of $\sim 126$~pc, \candtwo lies roughly 34 pc below the Galactic plane. Using the Gaia DR2 proper motion of $(\mu_\alpha,\mu_\delta)=(62.064 \pm 0.077, -48.245 \pm 0.051)~{\rm mas~yr}^{-1}$, the Gaia parallax, and the absolute radial velocity as determined from the TRES spectroscopy of $22.4 \pm 0.1~{\rm km~s^{-1}}$, we find
that \candthree has a three-dimensional Galactic space motion of $(U,V,W)= (-14.9 \pm 0.1, -34.5 \pm 0.3, 13.5 \pm 0.1)~{\rm km~s^{-1}}$, where again positive $U$ is in the direction of the Galactic center, and we have adopted the \citet{Coskunoglu:2011} determination of the solar motion with respect to the local standard of rest. These values yield a 98.3\% probability that \candthree is a thin disk star, according to the classification scheme of \citet{Bensby:2003}.

Note that stars of the mass of \candthree typically have spectral types of roughly G1V \citep{pecautmamajek13}, and in fact \candthree has a $T_{\rm eff}$ and colors that are roughly consistent with this spectral type \citep{pecautmamajek13}.  The radius and luminosity of \candthree are $R_*\sim 1.23~R_\odot$ and $L_*\sim 1.65~L_\odot$; again, these are roughly consistent, although slightly larger, than would be expected for a zero age main sequence star of its mass and spectral type \citep{pecautmamajek13}.  The Galactic velocities of \candthree are somewhat larger than typical thin disk stars.  Together, these pieces of information suggest that \candthree is likely a roughly solar-mass star, with an age that is somewhat larger than the average age of the Galactic thin disk, that is roughly 70\% of the way through its main-sequence lifetime.  Indeed, when combined with the estimate of its metallicity, we can roughly characterize \candthree as a slightly older, slightly more massive analog of the sun. 

As with \candtwo, this conclusion is corroborated by the properties of \candthree inferred from the global fit to the transit, radial velocity, SED, and parallax data described in \ref{sec:exofast}.  When comparing the \logg and \Teff from the global fit to MIST evolutionary tracks \citep{dotter16, choi16}, we infer that \candthree has an age of $\sim 7.1$~Gyr and is indeed just over halfway through its main sequence lifetime. 

\section{Planet Characterization} 
\label{sec:planet}

\subsection{EXOFASTv2 Global Fit}
\label{sec:exofast}
To determine the system parameters for both \candtwo and \candthree, we perform a simultaneous fit using exoplanet global fitting suite EXOFASTv2 \citep{Eastman:2017}. EXOFASTv2 is based largely on the original EXOFAST \citep{Eastman:2013} but is now more flexible and can, among many other features, simultaneously fit multiple RV instruments and the spectral energy distribution (SED) along with the transit data. Specifically, for each system we fit the flattened {\it K2} light curve, accounting for the long cadence smearing; the SED; and the radial velocity data. To constrain the stellar parameters, we used the MESA Isochrones \& Stellar Tracks \citep[MIST,][]{dotter16, choi16}, the broad band photometry, and the parallax from Gaia summarized in Table \ref{tab:stellar}. In addition, we set priors on \Teff and \feh \ from the Keck/HIRES spectra described in Section~\ref{spec} and enforced upper limits on the V-band extinction from the \citet{Schlegel1998} dust maps of 0.035 for \candtwo and 0.4765 for \candthree. We used the online EXOFAST tool\footnote{\url{http://astroutils.astronomy.ohio-state.edu/exofast/exofast.shtml}} to refine our starting values prior to the EXOFASTv2 fit.

We note that the fit yielded bimodal posterior distributions for the age and mass of \candtwo, with the age distribution showing peaks at 7.53 and 4.18 Gyr, and the mass peaking at $1.157 \ M_\odot$ and $1.324 \ M_\odot$. The two peaks correspond to two solutions with the star being a subgiant and a main sequence dwarf respectively. This degeneracy is also present when we repeat our global fits using the integrated Yale-Yonsei stellar tracks \citep{Yi:2001} instead of MIST. We also attempted an empirical fit using only the transits, RVs, SED, broadband photometry and Gaia parallax but no isochrones, and the resulting mass distribution, with error bars as large as 80\%, does not offer any useful insight. This degeneracy may be broken with better constraints on the eccentricity of the planet or asteroseismic analyses\footnote{In an independent discovery paper, \citet{vaneylen18} found a stellar mass consistent with the lower of the two masses through asteroseismology.}, but in this paper, we report both solutions agnostically. Even so, the resulting planet masses from these two solutions are consistent to within $1\sigma$ because the error on planet mass is dominated by the uncertainty on the RV semi-amplitude.

\begin{figure*}[bh]
\centerline{%
\includegraphics[width=0.5\textwidth, trim={2cm 12.5cm 8cm 8cm}, clip]{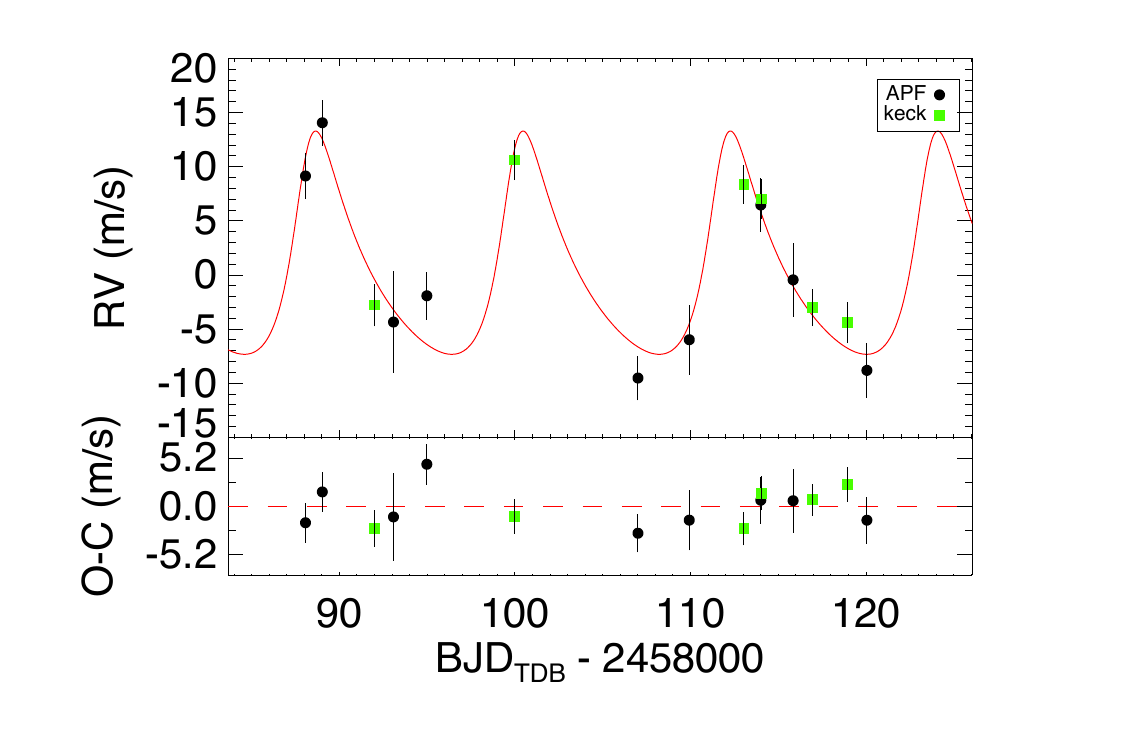}%
\includegraphics[width=0.5\textwidth, trim={2cm 12.5cm 8cm 8cm}, clip]{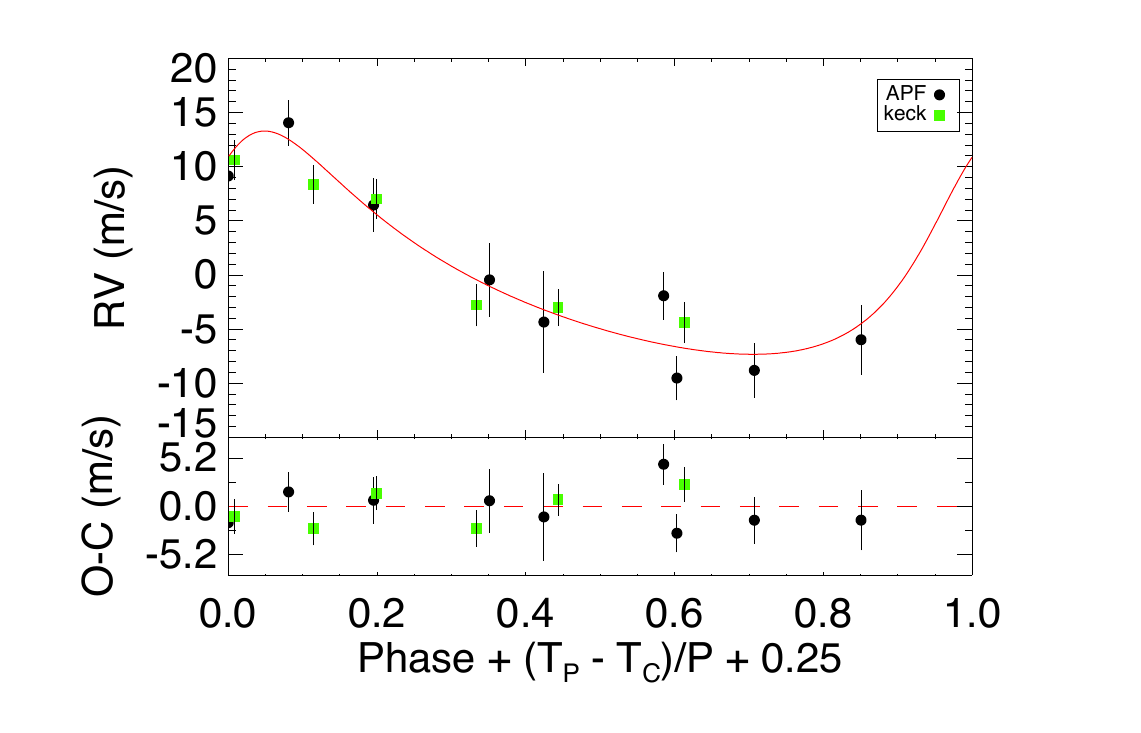}%
}%
\caption{{\it Left}: The RV time series of \candtwo. In each panel, the green squares are the HIRES data and black circles are the APF data. The maximum-likelihood eccentric Keplerian orbital model is plotted in red. The instrumental offset has been subtracted from each data set and the model. The uncertainties plotted include the RV jitter terms listed in Table~\ref{tbl:Parameters} added in quadrature with the measurement uncertainties for all RVs. Below are the residuals to the maximum-likelihood eccentric orbit model. {\it Right}: same as the left panel, but phase-folded to the best-fit ephemeris. The X-axis is defined such that the primary transit occurs at 0.25, where $T_P$ is the time of periastron, $T_C$ is the time of transit, and $P$ is the period.}
\label{fig:rv2} 
\end{figure*}

\begin{figure*}[htb]
\centerline{%
\includegraphics[width=0.5\textwidth, trim={2cm 12.5cm 8cm 8cm}, clip]{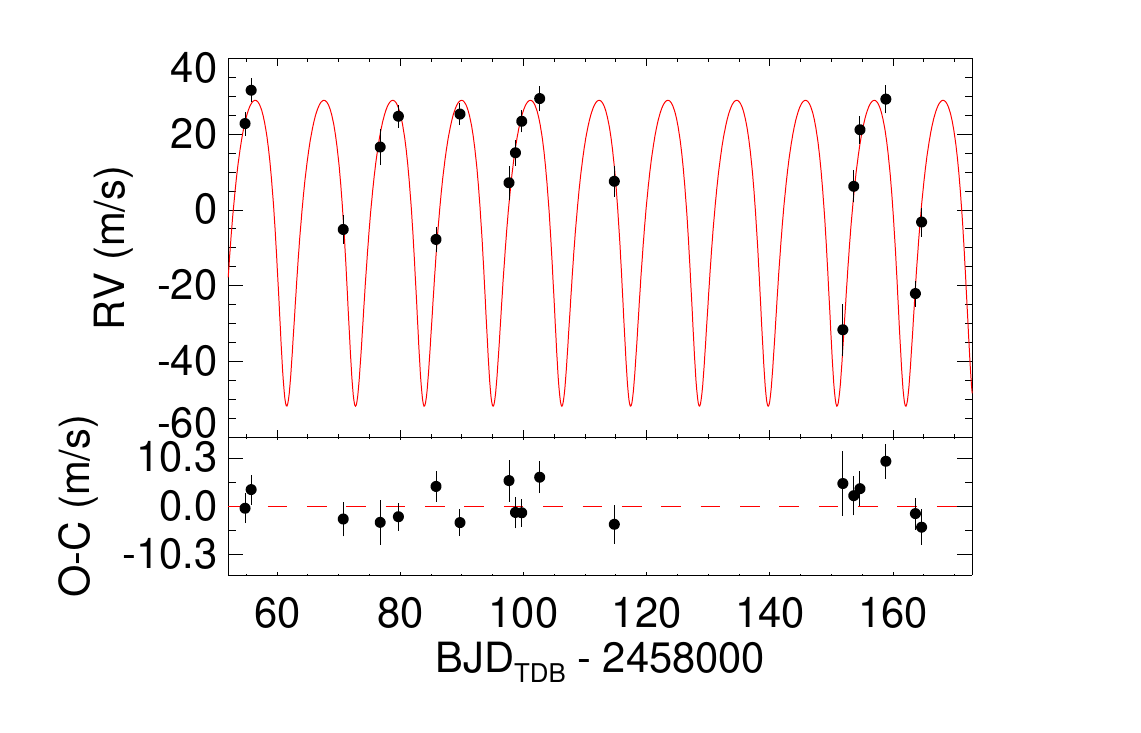}%
\includegraphics[width=0.5\textwidth, trim={2cm 12.5cm 8cm 8cm}, clip]{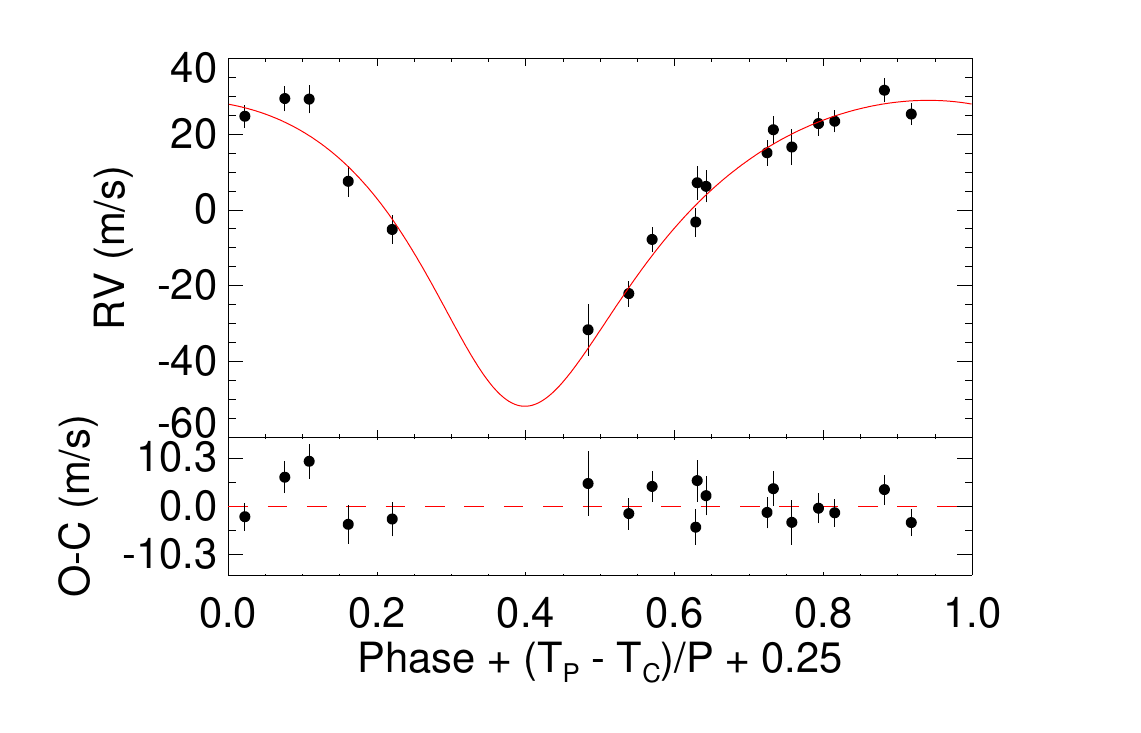}%
}%
\caption{Same as Fig.~\ref{fig:rv2}, but for \candthree.}
\label{fig:rv3} 
\end{figure*}

\begin{figure*}[htb]
\centerline{%
\includegraphics[width=0.5\textwidth, trim={0cm 6.5cm 0cm 6.5cm}, clip]{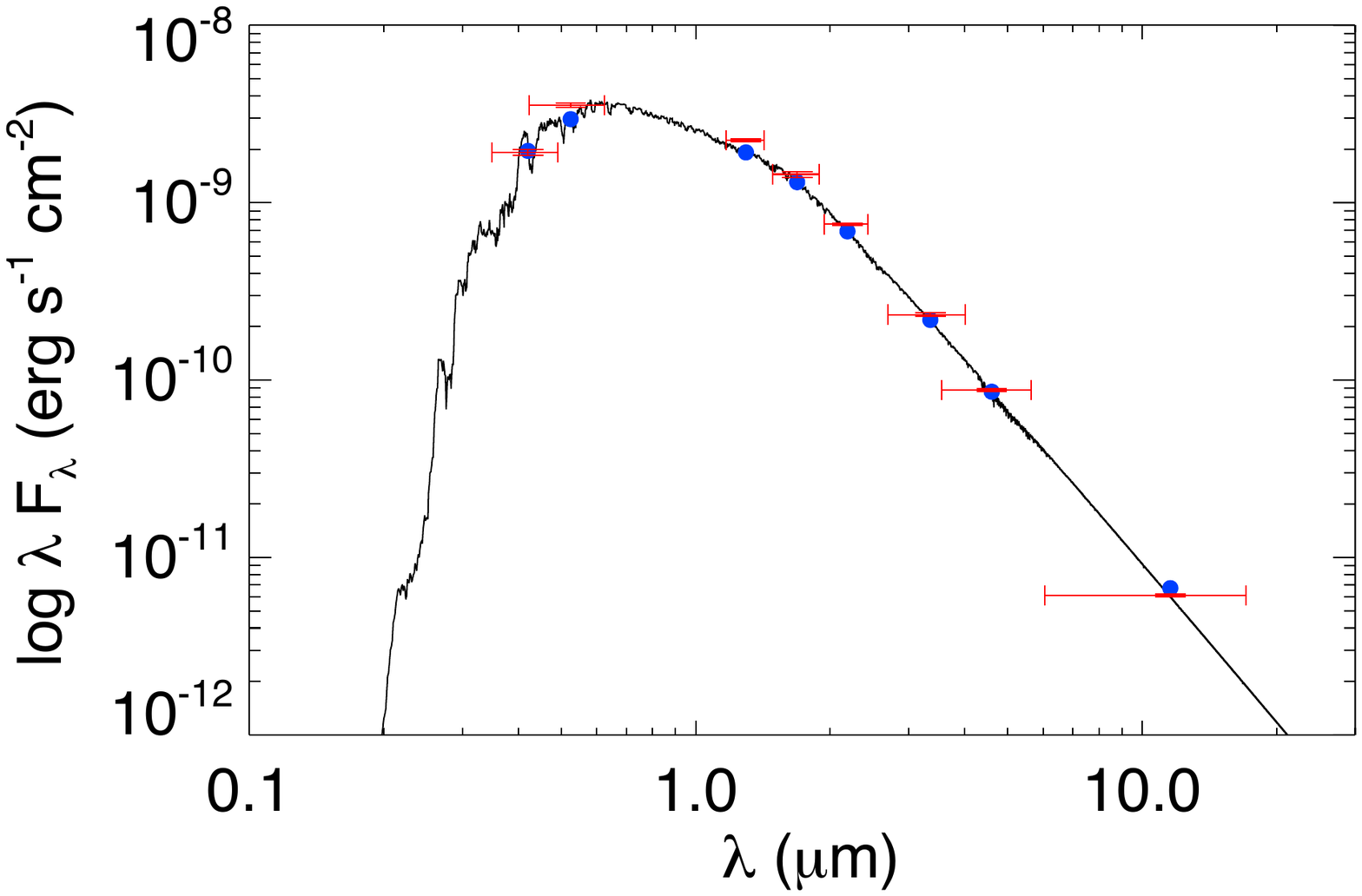}%
\includegraphics[width=0.5\textwidth, trim={0cm 6.5cm 0cm 6.5cm}, clip]{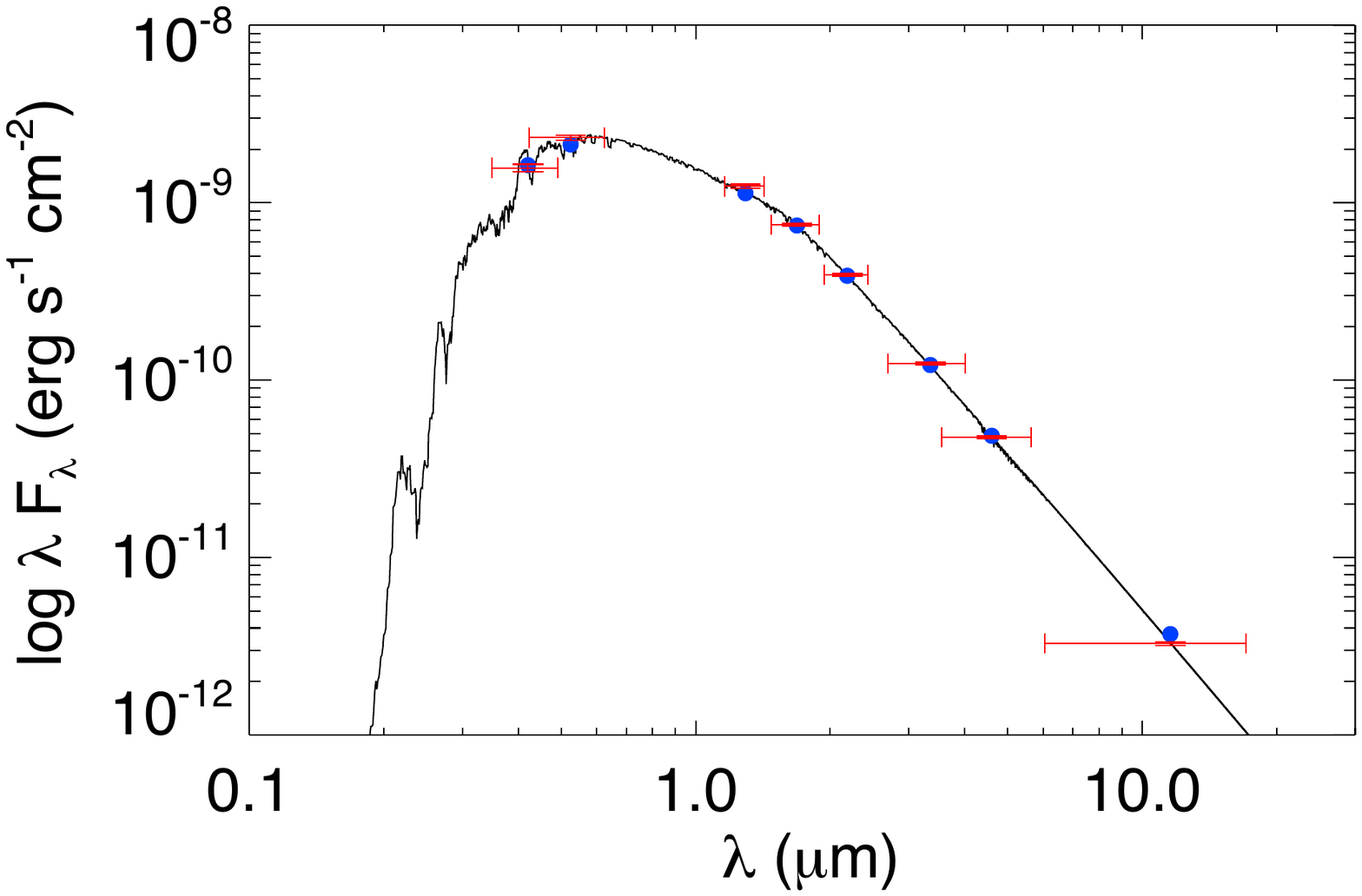}%
}%
\caption{SED fits to \candtwo (left) and \candthree (right) from EXOFASTv2. The red points show observed values, with the vertical error bars representing $1\sigma$ measurement uncertainties and horizontal error bars representing the widths of the bandpasses. The blue points are the model fluxes in the observed bandpasses. The solid lines show the model fits.}
\label{fig:sed} 
\end{figure*}

\begin{figure*}[htb]
\centering
\begin{minipage}{.49\linewidth}
\centering
\subfloat[]{\label{main:a}
\includegraphics[width=\textwidth, trim={3cm 2cm 2cm 2cm}, clip]{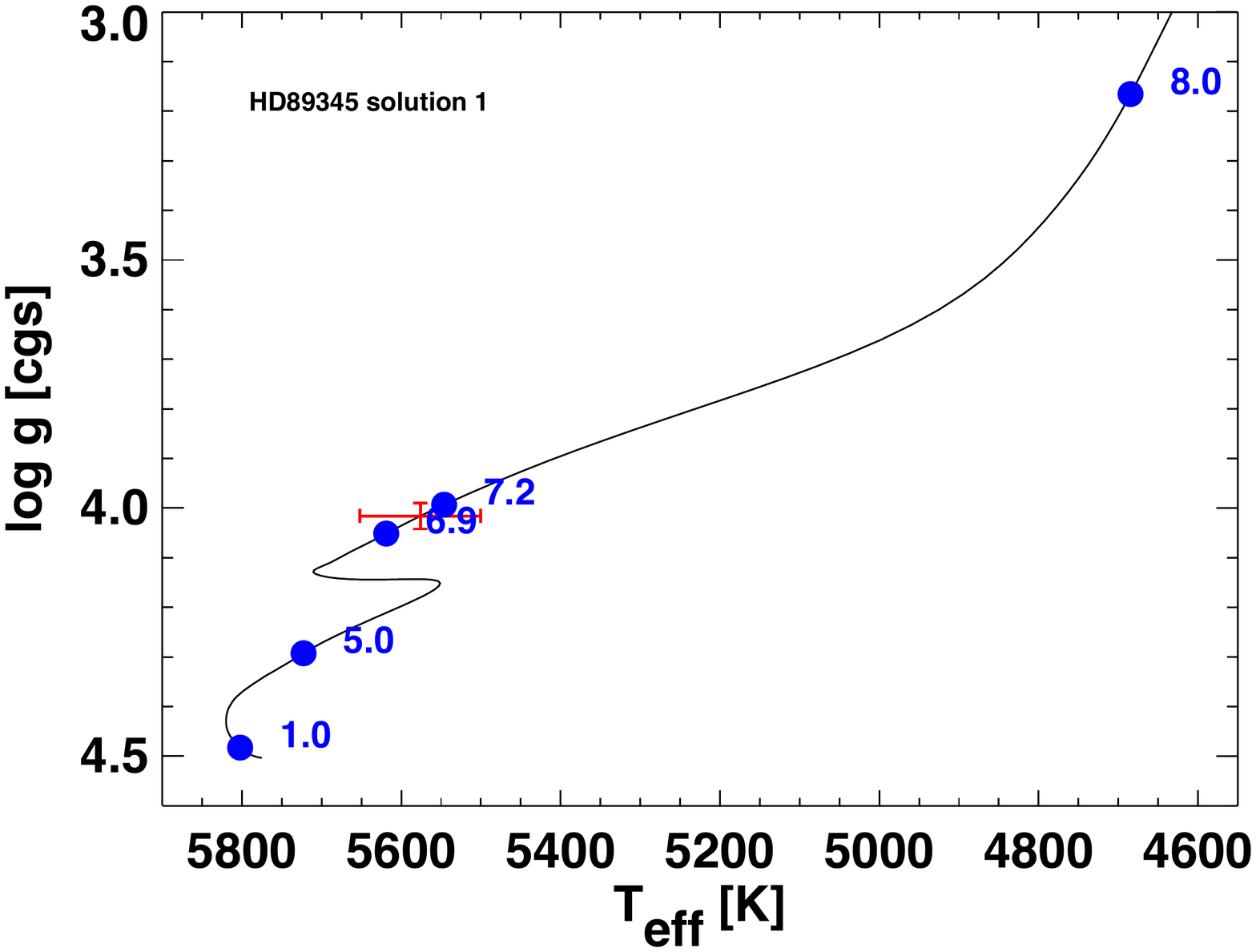}}
\end{minipage}%
\begin{minipage}{.49\linewidth}
\centering
\subfloat[]{\label{main:b}
\includegraphics[width=\textwidth, trim={3cm 2cm 2cm 2cm}, clip]{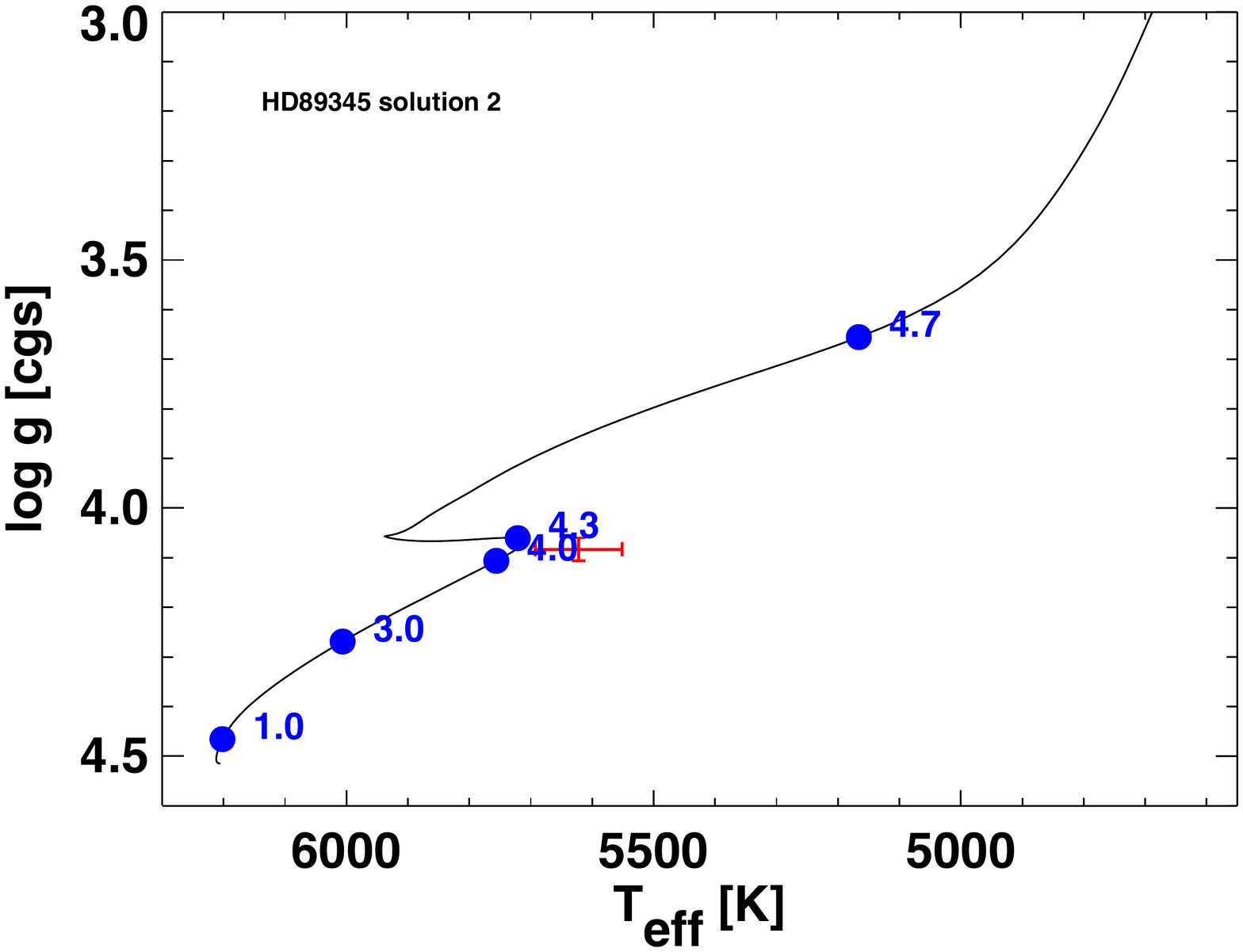}}
\end{minipage}\par\medskip
\centering
\subfloat[]{\label{main:c}
\includegraphics[width=0.49\textwidth, trim={3cm 2cm 2cm 2cm}, clip]{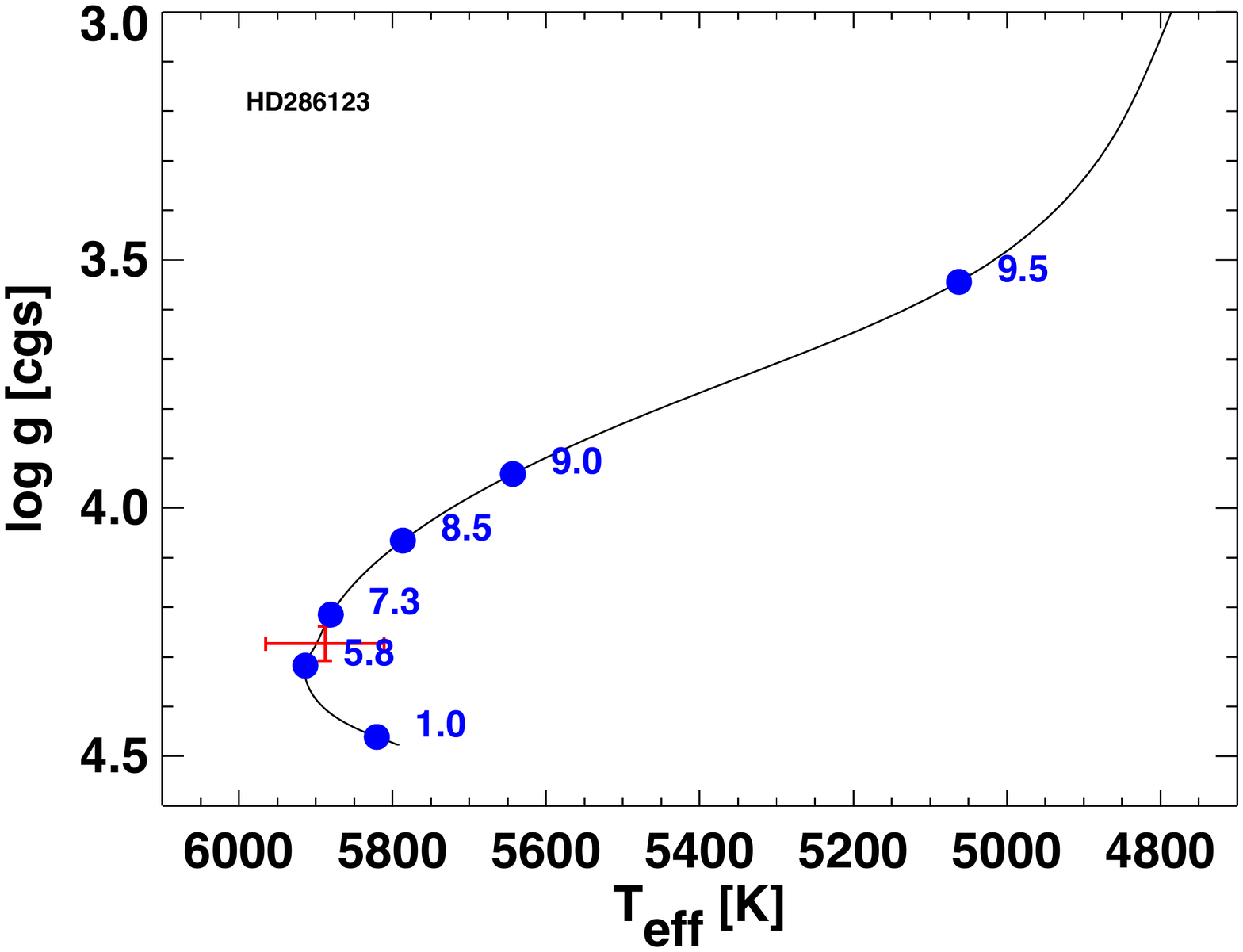}}
\caption{The locations of \candtwo solution 1 (a), \candtwo solution 2 (b) and \candthree (c) in the Kiel diagram. The median \Teff and \logg from the global model fit are shown as red points, while the black lines show MIST evolutionary tracks for stars with the best-fit values of \Mstar \ and [Fe/H]; the locations on the best-fit model corresponding to several values of stellar age are shown as blue points, with ages quoted in Gyr. The red points do not fall exactly on the evolutionary tracks at the median ages quoted in Table~\ref{tbl:Parameters}, because the median values in Table~\ref{tbl:Parameters} are drawn from individual posterior distributions and are not always exactly self-consistent.}
\label{fig:yy} 
\end{figure*}

See Fig.~\ref{fig:lcs} for the final transit fits, Figs.~\ref{fig:rv2} \& \ref{fig:rv3} for the final RV fits, Fig.~\ref{fig:sed} for the final SED fit from our EXOFASTv2 global fit, and Fig.~\ref{fig:yy} for the best-fit evolutionary tracks. The median values of the posterior distributions of the system parameters are shown in Table \ref{tbl:Parameters}.

{\renewcommand\normalsize{\tiny}%
\normalsize\begin{table*}
 \scriptsize
\centering
\setlength\tabcolsep{1.5pt}
\caption{Median values and 68\% confidence intervals for the physical and orbital parameters of the \candtwo and \candthree systems}
\label{tbl:Parameters}
  \begin{tabular}{lcccc}
  \hline
  \hline
   Parameter & Units & \multicolumn{2}{c}\candtwo  & \candthree\\
 & & Solution 1 (subgiant) & Solution 2 (main sequence) & \\
 \hline
Stellar Parameters & & & \\
\hline
~~~~$M_*$\dotfill &Mass (\msun)\dotfill &   $1.157^{+0.040}_{-0.045}$     &$1.324^{+0.044}_{-0.041}$        &$1.039^{+0.071}_{-0.065}$\\
~~~~$R_*$\dotfill &Radius (\rsun)\dotfill &   $1.747^{+0.049}_{-0.050}$     &$1.733\pm0.047$        &$1.233^{+0.026}_{-0.025}$\\
~~~~$L_*$\dotfill &Luminosity (\lsun)\dotfill &$2.66^{+0.15}_{-0.16}$         &$2.71\pm0.15$     &$1.646^{+0.080}_{-0.079}$\\
~~~~$\rho_*$\dotfill &Density (cgs)\dotfill & $0.305^{+0.027}_{-0.025}$     &$0.359^{+0.029}_{-0.026}$        &$0.782^{+0.077}_{-0.070}$\\
~~~~$\log{g}$\dotfill &Surface gravity (cgs)\dotfill &$4.016\pm0.026$        &$4.083^{+0.023}_{-0.022}$     &$4.273\pm0.035$\\
~~~~$T_{\rm eff}$\dotfill &Effective Temperature (K)\dotfill &$5576^{+73}_{-76}$ &$5622^{+70}_{-71}$               &$5888^{+71}_{-77}$\\
~~~~$[{\rm Fe/H}]$\dotfill &Metallicity \dotfill &$0.421^{+0.046}_{-0.054}$ & $0.436^{+0.040}_{-0.050}$              &$0.051^{+0.058}_{-0.056}$\\
~~~~$Age$\dotfill &Age (Gyr)\dotfill &         $7.53^{+1.3}_{-0.99}$      &  $4.18^{+0.64}_{-0.69}$   &$7.1^{+3.1}_{-2.7}$\\ 
~~~~$A_v$\dotfill &V-band extinction \dotfill &$0.017^{+0.012}_{-0.011}$ &$0.017\pm0.012$               &$0.017^{+0.012}_{-0.011}$\\
~~~~$\sigma_{SED}$\dotfill &SED photometry error scaling \dotfill & $5.4^{+2.2}_{-1.3}$ &     $5.2^{+2.1}_{-1.3}$         &$3.28^{+1.3}_{-0.82}$\\
~~~~$d$\dotfill &Distance (pc)\dotfill &        $132.8^{+1.8}_{-1.7}$       &$132.7\pm1.7$ & $131.7\pm1.7$ \\
~~~~$\pi$\dotfill &Parallax (mas)\dotfill &    $7.533\pm0.098$            &$7.533^{+0.098}_{-0.097}$ & $7.594^{+0.10}_{-0.098}$\\
\hline
 Planet Parameters & & &  \\
\hline
~~~~$P$\dotfill &Period (days)\dotfill & $11.81430\pm0.00020$        &$11.81430^{+0.00020}_{-0.00019}$       &$11.168459\pm0.000017$\\
~~~~$R_P$\dotfill &Radius (\rj)\dotfill &$0.660^{+0.028}_{-0.030}$   &$0.648^{+0.029}_{-0.028}$             &$1.058^{+0.023}_{-0.022}$\\
~~~~$T_C$\dotfill &Time of Transit (\bjdtdb)\dotfill &$2457913.8052^{+0.0011}_{-0.0010}$        &$2457913.80504^{+0.0011}_{-0.00094}$       &$2457858.856812^{+0.000042}_{-0.000046}$\\
~~~~$a$\dotfill &Semi-major axis (AU)\dotfill &$0.1066^{+0.0012}_{-0.0014}$      &$0.1115\pm0.0012$          &$0.0991^{+0.0022}_{-0.0021}$\\
~~~~$i$\dotfill &Inclination (Degrees)\dotfill &$87.21^{+0.43}_{-0.22}$                &$87.56^{+0.59}_{-0.24}$ & $89.61^{+0.26}_{-0.29}$\\
~~~~$e$\dotfill &Eccentricity \dotfill &   $0.220^{+0.095}_{-0.13}$      &$0.22^{+0.10}_{-0.12}$      &$0.255^{+0.034}_{-0.036}$\\
~~~~$\omega_*$\dotfill &Argument of Periastron (Degrees)\dotfill &$-13^{+58}_{-27}$        &$-16^{+56}_{-27}$        &$170.9^{+5.7}_{-340}$\\
~~~~$T_{eq}$\dotfill &Equilibrium temperature (K)\dotfill &$1089^{+15}_{-16}$        &$1068^{+14}_{-15}$        &$1001\pm14$\\
~~~~$M_P$\dotfill &Mass (\mj)\dotfill & $0.110^{+0.017}_{-0.018}$    &$0.121^{+0.018}_{-0.019}$           &$0.387^{+0.044}_{-0.042}$\\
~~~~$K$\dotfill &RV semi-amplitude (m/s)\dotfill &$9.2\pm1.5$          &$9.2\pm1.5$       &$35.4^{+4.6}_{-4.3}$\\
~~~~$logK$\dotfill &Log of RV semi-amplitude \dotfill &   $0.962^{+0.064}_{-0.080}$          	&$0.965^{+0.064}_{-0.078}$	   &$1.550^{+0.053}_{-0.056}$\\
~~~~$R_P/R_*$\dotfill &Radius of planet in stellar radii \dotfill &$0.0389^{+0.0011}_{-0.0012}$  &$0.0384\pm0.0012$    	       &$0.08811^{+0.00031}_{-0.00017}$\\
~~~~$a/R_*$\dotfill &Semi-major axis in stellar radii \dotfill &$13.11^{+0.38}_{-0.36}$          &$13.84^{+0.36}_{-0.34}$		     &$17.28^{+0.55}_{-0.53}$\\
~~~~$\delta$\dotfill &Transit depth (fraction)\dotfill &        $0.001509^{+0.000085}_{-0.000091}$ & $0.001477^{+0.000094}_{-0.000090}$   			   &$0.007764^{+0.000054}_{-0.000030}$\\
~~~~$Depth$\dotfill &Flux decrement at mid transit \dotfill &$0.001509^{+0.000085}_{-0.000091}$ & $0.001477^{+0.000094}_{-0.000090}$               &$0.007764^{+0.000054}_{-0.000030}$\\
~~~~$\tau$\dotfill &Ingress/egress transit duration (days) \dotfill &$0.0149^{+0.0045}_{-0.0037}$ &$0.0135^{+0.0046}_{-0.0034}$        	      &$0.01717^{+0.00047}_{-0.00020}$\\
~~~~$T_{14}$\dotfill &Total transit duration (days)\dotfill & $0.2389^{+0.0037}_{-0.0033}$         &$0.2378^{+0.0038}_{-0.0030}$   		  &$0.20959^{+0.00033}_{-0.00028}$\\
~~~~$T_{FWHM}$\dotfill &FWHM transit duration (days)\dotfill &$0.2239\pm0.0016$        &$0.2241\pm0.0016$    		    &$0.19233^{+0.00026}_{-0.00027}$\\
~~~~$b$\dotfill &Transit Impact parameter \dotfill &  $0.645^{+0.089}_{-0.15}$          		&$0.60^{+0.11}_{-0.19}$		    &$0.108^{+0.080}_{-0.072}$\\
~~~~$b_S$\dotfill &Eclipse impact parameter \dotfill &$0.568^{+0.057}_{-0.079}$              &$0.509^{+0.051}_{-0.077}$ 			&$0.114^{+0.082}_{-0.076}$\\
~~~~$\tau_S$\dotfill &Ingress/egress eclipse duration (days)\dotfill &$0.0129^{+0.0018}_{-0.0020}$ &$0.0114^{+0.0013}_{-0.0016}$            &$0.0182^{+0.0011}_{-0.0010}$\\
~~~~$T_{S,14}$\dotfill &Total eclipse duration (days)\dotfill &$0.232^{+0.030}_{-0.021}$   &$0.226^{+0.039}_{-0.026}$           		 &$0.221^{+0.013}_{-0.012}$\\
~~~~$T_{S,FWHM}$\dotfill &FWHM eclipse duration (days)\dotfill &$0.219^{+0.029}_{-0.018}$ &$0.215^{+0.038}_{-0.024}$         	  &$0.202^{+0.012}_{-0.011}$\\
~~~~$\delta_{S,3.6\mu m}$\dotfill &Blackbody eclipse depth at 3.6$\mu$m (ppm)\dotfill &$40.0^{+3.1}_{-3.3}$       &$36.0\pm2.9$        &$136.9^{+7.6}_{-7.4}$\\
~~~~$\delta_{S,4.5\mu m}$\dotfill &Blackbody eclipse depth at 4.5$\mu$m (ppm)\dotfill &$65.3^{+4.7}_{-5.1}$       &$59.7^{+4.6}_{-4.5}$         &$239^{+11}_{-10.}$\\
~~~~$\rho_P$\dotfill &Density (cgs)\dotfill &   $0.471^{+0.094}_{-0.084}$         		&$0.547^{+0.11}_{-0.096}$	    &$0.405^{+0.046}_{-0.044}$\\
~~~~$logg_P$\dotfill &Surface gravity \dotfill &$2.795^{+0.068}_{-0.079}$         		&$2.852^{+0.066}_{-0.076}$	   &$2.933^{+0.045}_{-0.048}$\\
~~~~$\fave$\dotfill &Incident Flux (\fluxcgs)\dotfill &$0.303^{+0.021}_{-0.022}$         	&$0.280^{+0.019}_{-0.020}$	     &$0.214\pm0.011$\\
~~~~$T_P$\dotfill &Time of Periastron (\bjdtdb)\dotfill &$2457911.30^{+1.5}_{-0.89}$ &$2457911.23^{+1.4}_{-0.89}$             &$2457860.59^{+0.24}_{-0.20}$\\
~~~~$T_S$\dotfill &Time of eclipse (\bjdtdb)\dotfill &$2457909.29^{+0.65}_{-1.1}$     &$2457909.33^{+0.63}_{-1.1}$          &$2457862.67^{+0.25}_{-0.23}$\\
~~~~$T_A$\dotfill &Time of Ascending Node (\bjdtdb)\dotfill &$2457911.42^{+0.36}_{-0.43}$       &$2457911.40^{+0.35}_{-0.42}$      &$2457855.32\pm0.17$\\
~~~~$T_D$\dotfill &Time of Descending Node (\bjdtdb)\dotfill &$2457917.54^{+1.0}_{-0.84}$      &$2457917.62^{+1.1}_{-0.88}$         &$2457860.70^{+0.15}_{-0.14}$\\
~~~~$ecos{\omega_*}$\dotfill & \dotfill &  $0.185^{+0.087}_{-0.15}$        &$0.189^{+0.084}_{-0.15}$      &$-0.252^{+0.036}_{-0.033}$\\
~~~~$esin{\omega_*}$\dotfill & \dotfill &   $-0.03^{+0.12}_{-0.13}$        &   $-0.04^{+0.12}_{-0.14}$ &$0.026\pm0.029$\\
~~~~$M_P\sin i$\dotfill &Minimum mass (\mj)\dotfill &$0.110^{+0.017}_{-0.018}$           &$0.121^{+0.018}_{-0.019}$    &$0.387^{+0.044}_{-0.042}$\\
~~~~$M_P/M_*$\dotfill &Mass ratio \dotfill &$0.000091^{+0.000014}_{-0.000015}$      &  $0.000087^{+0.000013}_{-0.000014}$     &$0.000355^{+0.000047}_{-0.000044}$\\
~~~~$d/R_*$\dotfill &Separation at mid transit \dotfill &$13.1^{+1.4}_{-1.5}$            		&$13.9^{+1.5}_{-1.6}$	 &$15.8^{+1.1}_{-1.0}$\\
\hline
 Wavelength Parameters & Kepler & &  \\
\hline
~~~~$u_{1, Kepler}$\dotfill &linear limb-darkening coeff \dotfill &  $0.437\pm0.038$      		    &$0.432^{+0.037}_{-0.038}$    &$0.412\pm0.011$\\
~~~~$u_{2, Kepler}$\dotfill &quadratic limb-darkening coeff \dotfill &$0.206^{+0.046}_{-0.045}$     &$0.214\pm0.046$         &$0.211\pm0.027$\\
\hline
 Telescope Parameters & Kepler & &  \\
\hline
~~~~$\gamma$\dotfill &APF instrumental offset (m/s)\dotfill &  $0.2^{+1.8}_{-1.7}$     & $0.2\pm1.7$ &$-11.8\pm2.2$\\
~~~~$\gamma$\dotfill &HIRES instrumental offset (m/s)\dotfill & $-3.0^{+1.7}_{-1.6}$  &$-3.1^{+1.7}_{-1.6}$  &---\\
~~~~$\sigma_J$\dotfill &APF RV jitter \dotfill & $4.2^{+2.9}_{-2.2}$   &$4.1^{+2.9}_{-2.2}$   &$3.7^{+1.6}_{-1.4}$\\
~~~~$\sigma_J$\dotfill &HIRES RV jitter \dotfill &  $4.1^{+1.5}_{-1.1}$      &$4.0^{+1.5}_{-1.1}$  &---\\
~~~~$\sigma_J^2$\dotfill &APF RV jitter variance \dotfill & $17^{+33}_{-14}$     &$16^{+33}_{-13}$   &$13.8^{+15}_{-8.4}$\\
~~~~$\sigma_J^2$\dotfill &HIRES RV jitter variance \dotfill &  $16.4^{+15}_{-7.5}$ &$16.3^{+14}_{-7.4}$    &---\\
\hline
 Transit Parameters & Kepler & &  \\
\hline
~~~~$\sigma^{2}$\dotfill &Added Variance \dotfill & $0.00000000007^{+0.00000000050}_{-0.00000000045}$    &$0.00000000006^{+0.00000000051}_{-0.00000000045}$ &$0.00000000038^{+0.00000000015}_{-0.00000000013}$\\
~~~~$F_0$\dotfill &Baseline flux \dotfill & $0.9999984\pm0.0000050$       &$0.9999983\pm0.0000051$ &$0.9999994\pm0.0000027$\\
\hline
 \end{tabular}
\end{table*}

}

\subsection{RV Analysis with \texttt{RadVel}} 
For comparison with EXOFASTv2, we also analyze the RV time series using another widely used, publicly available RV fitting package \texttt{RadVel}\footnote{\url{http://radvel.readthedocs.io/en/master/index.html}} \citep{radvel}. We impose a Gaussian prior on the orbital period and times of conjunction of \candtwo and \candthree with means and standard deviations derived from transit photometry and given in Table~\ref{tbl:Parameters}. We initially included a constant radial acceleration term, $dv/dt$, but the result is consistent with zero for both systems. Therefore we fix $dv/dt$ to zero. The remaining free parameters are the velocity semi-amplitudes, the zero-point offsets for each instrument, and the jitter terms for each instrument. The jitter terms are defined in Equation 2 of \citet{fulton15} and serve to capture the stellar jitter and instrument systematics such that the reduced $\chi^2$ of the best-fit model is close to 1. To calculate $M_P \sin i$, we adopt the median stellar masses in Table~\ref{tbl:Parameters} and their quoted error bars.

The fitting procedure is identical to that described in \citet{sinukoff16}. The best-fit Keplerian orbital solutions are in agreement with those from EXOFASTv2 at the $1\sigma$ level. 


\section{Discussion}
\label{sec:conclusions}
\subsection{Potential for Atmospheric Characterization}
Sub-Jovian gas giants are particularly interesting targets for atmospheric studies because a wide range of atmospheric compositions are possible. Yet the atmospheres of such planets, especially those more massive than Neptune but less massive than Saturn, have not been thoroughly studied, both because the host stars of most such systems are too faint for atmospheric characterization, and because the mass regime of sub-Saturns is relatively unpopulated. The two systems presented in this paper are therefore important additions to the small sample of sub-Jovian gas giants amenable to atmospheric characterization. 

\begin{figure*}[htb]
\centering
\includegraphics[width=\textwidth]{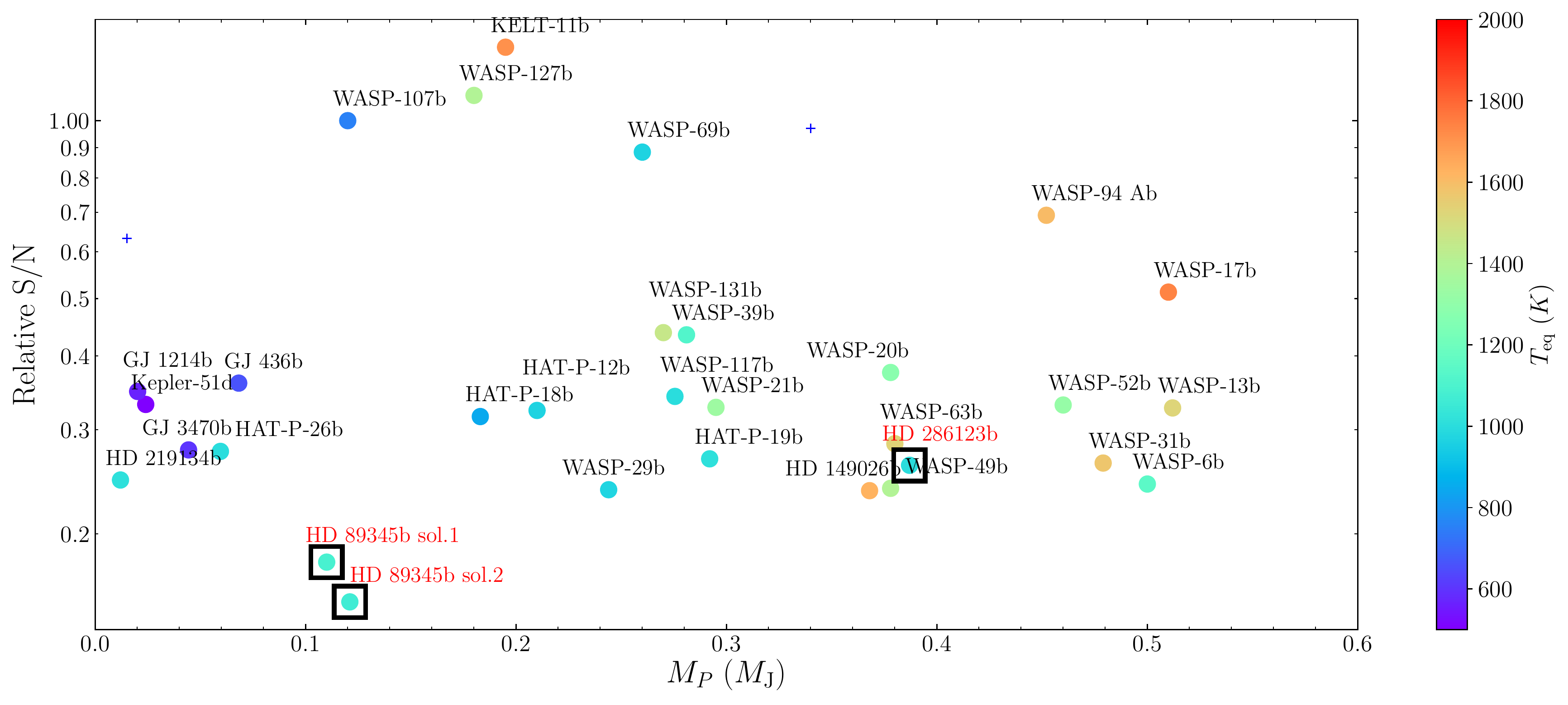}
\caption{Estimated SNR per transit for transmission spectroscopy, relative to that of WASP-107b, as a function of planetary mass for planets with $0.01\ M_\mathrm{J}<M_p<0.5\ M_\mathrm{J}$. Both solutions are shown for \pltwo. Small plus symbols denote planets with uncertain mass and/or radius measurements (error $>20\%$). Data retrieved from the NASA Exoplanet Archive on May 21, 2018.}
\label{fig:trans} 
\end{figure*}

With their bright host stars and low planet densities, both systems are promising targets for transit transmission spectroscopy. Such observations could provide insight into the planets' bulk composition and formation histories by measuring the elemental composition of their atmospheres, and overall metal enrichment. We calculated the expected SNR per transit compared to the expected scale height of each planet's atmosphere, and compared the results with other known transiting planets with $0.01\ M_\mathrm{J}<M_p<0.5\ M_\mathrm{J}$. Specifically, we calculated the SNR as 
\begin{gather}
    \mathrm{SNR} \propto \frac{R_pH \sqrt{Ft_{14}}}{R_{\star}^2} \\
    H = \frac{k_\mathrm{b}T_\mathrm{eq}}{\mu g}
\end{gather}
where \Rp is the planet’s radius, \Rstar is the star's radius, $H$ is the planet atmosphere's scale height, $k_\mathrm{b}$ is Boltzmann's constant, \Teq is the planet's equilibrium temperature, $\mu$ is the atmosphere's mean molecular weight, $g$ is the planet's surface gravity, $t_{14}$ is the transit duration, and $F$ is the flux from the star. To simplify the comparison, we assumed the planets' atmospheres were dominated by molecular hydrogen and $\mu=2$ for all cases. We also calculated $F$ from the host stars' $H$-band magnitudes to test suitability for observations with the Hubble Space Telescope's Wide Field Camera 3 instrument. Fig.~\ref{fig:trans} shows the expected SNR for transmission spectroscopy (normalized such that the predicted SNR for WASP-107b is unity) against planet masses for \pltwo, \plthree, and 30 known planets with the highest estimated SNR. For reference, \citet{kreidberg17} detected water features at $6.5 \sigma$ confidence with a single HST/WFC3 transit observation of WASP-107b, the benchmark for comparison. \plthree appears to be one of the coolest Saturn-sized planets that are amenable to transmission spectroscopy. Notably, many known planets with the highest expected SNR, including GJ 1214b \citep{kreidberg14}, GJ 3470b \citep{ehrenreich14} and GJ 436b \citep{knutson14a},
were found to show essentially featureless transmission spectra, indicating the existence of hazes, clouds, or atmospheres with high molecular weight. So the estimated SNR does not necessarily mean that we will detect spectral features in the atmospheres of \pltwo and \plthree. Nevertheless, not all such planets have featureless spectra \citep{crossfieldkreidberg}. Past works have found that a planet's likelihood of being cloudy/hazy is correlated with its equilibrium temperature: at temperatures below roughly 1000 K, methane is abundant and can easily photolyze to produce hydrocarbon hazes \citep[e.g.][]{fortney13, morley13}. These predictions are borne out in observations of transmission spectra showing that hotter planets tend to have larger spectral features \citep[e.g.][]{stevenson16, crossfieldkreidberg, fu17}. At \Teq $\approx 1000$ K, \pltwo and \plthree are less likely to be hazy and there are fewer condensible cloud species. It is therefore scientifically compelling to pursue transmission spectroscopy for these planets, both to increase the small sample of Neptune- to Saturn-sized planets with well-characterized atmospheres and to inform the choice of which {\it TESS} planets to observe to efficiently study the atmospheric composition of sub-Jovian planets.

In addition to transit spectroscopy, \plthree is also a good candidate for secondary eclipse detection. Table~\ref{tbl:Parameters} shows the blackbody eclipse depths at 3.6 $\mu$m and 4.5 $\mu$m, derived using the planet's equilibrium temperature assuming perfect redistribution and zero albedo, to test the feasibility of secondary eclipse observations with {\it Spitzer}. \plthree probes a different period, mass and temperature range from most other planets with secondary eclipse detections, and is one of the few targets that are good candidates for both transmission spectroscopy and secondary eclipse observations.

\subsection{The Evolutionary History of Close-in Giant Planets}
Both planets fall in the same period range as warm Jupiters, giant planets with incident irradiation levels near or below $2\times 10^8$ erg s$^{-1}$ cm$^{-2}$, corresponding to orbital periods longer than 10 days around Sun-like stars \citep{shporer17}. Like hot Jupiters, they may have formed {\it in situ}, or migrated inward through high eccentricity migration or disk migration. But at wider orbital separations than hot Jupiters, the orbits of warm Jupiters are less likely to be perturbed by tides raised on the star, and their eccentricity and stellar obliquity distributions may serve as the primordial (after emplacement) distributions for hot Jupiters. Previous works have found that the eccentricity distribution of warm Jupiters contains a low eccentricity component and a component with an approximately uniform distribution \citep{petrovichtremaine}. The former component cannot be easily explained by the high eccentricity tidal migration hypothesis, and the latter is a challenge for {\it in situ} formation or disk migration. This suggests that perhaps there is more than one migration mechanism at work. 

\begin{figure*}[htb]
\centering
\includegraphics[width=0.75\textwidth]{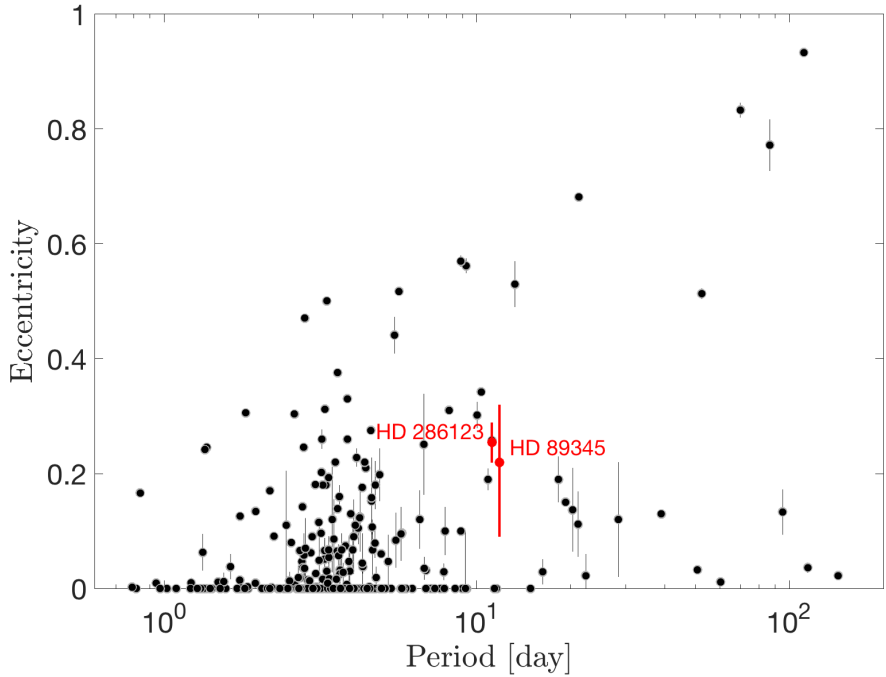}
\caption{Orbital eccentricity versus the log of the orbital period for transiting planets. The two new planets described in this paper are labeled and marked in red.
Data retrieved from the NASA Exoplanet Archive on April 20, 2018.}
\label{fig:perecc} 
\end{figure*}

Fig.~\ref{fig:perecc} shows \pltwo and \plthree in a period-eccentricity diagram along with other known planets. \plthree has a moderately high eccentricity compared to planets at similar periods. The eccentricity of \pltwo is only weakly constrained and driven away from zero largely by one data point. Given that, and the Lucy-Sweeney bias that tends to overestimate eccentricity due to the boundary at $e=0$ \citep{Lucy:1971}, we cannot consider the eccentricity of \pltwo to be significant without additional RV measurements. If these planets arrived at their present locations via high eccentricity migration, they must each be accompanied by a strong enough perturber to overcome precession caused by general relativity \citep{dong14}. Moreover, \citet{dong14} predicted that for warm Jupiters with orbital distances of 0.1-0.5 AU, the perturbers must have separations of $\sim$1.5-10 AU (period 2-30 years). Although we detected no significant linear trend in the RVs of \candtwo or \candthree, long-term RV monitoring may be able to reveal the existence of any distant companions.

Both planets are also favorable targets for stellar obliquity measurements. Among hot Jupiter systems, spin-orbit misalignment is more commonly seen among hot stars \citep[\Teff $\geq 6100$ K;][]{schlaufman10, winn10, albrecht12}, and among the cooler stars, those hosting misaligned hot Jupiters are all in the zone $a_\mathrm{min}/R_{\star} \geq 8$ \citep{daiwinn}. Hot Jupiters also tend to be more misaligned at longer orbital periods \citep{liwinn}. These observations have been construed as evidence for tidal realignment at work, but tidal realignment suffers from problems pointed out by \citet{mazeh15}, who found that the hot/cool obliquity distinction persists even in cases where tidal interactions should be negligible. The interpretation of warm Jupiters' stellar obliquities remains an outstanding problem. Resolving this problem requires a larger observational sample size, yet the set of warm Jupiters (and smaller planets) currently available for obliquity studies is very small. Both planets in this paper have $a/R_{\star}$ values beyond the threshold for alignment found by \citet{daiwinn}, and the tidal effects on them are expected to be relatively weak. Measuring their stellar obliquities can potentially offer insight into their migration history and tidal realignment theories. One possible method is to measure the Rossiter-McLaughlin (RM) effect, whose maximum semi-amplitude is approximately
\begin{equation}
    \Delta V_\mathrm{RM} \approx \Big( \frac{R_P}{\Rstar} \Big)^2 \sqrt{1-b^2}(\vsini)
\end{equation}
where $b$ is the impact parameter and \vsini is the projected equatorial rotation velocity of the star. Substituting values in Tables~\ref{tab:stellar} and~\ref{tbl:Parameters}, we obtain $V_\mathrm{RM}\approx 4$ \ms for \pltwo and $V_\mathrm{RM}\approx 23$ \ms for \plthree. Both should be detectable by modern spectrographs .

\subsection{Constraining Planet Inflation Models}
\label{sec:inflate}
Many of the proposed mechanisms for explaining the inflated radii of giant planets are related to the irradiation the planet receives from its host star \citep[c.f.][]{burrows07, fortney07}. The relation to irradiation seems to be empirically confirmed. For example, radius enhancement is common if the planet receives at least $\sim 2\times10^8\,\mathrm{erg \ s^{-1} \ cm^{-2}}$, and mostly absent below that threshold \citep{millerfortney, demoryseager}, and \citet{hartman16} argued that planets appear to re-inflate when their stars increase in luminosity as they leave the main sequence.

\begin{figure*}[hbt]
\centering
\includegraphics[width=0.75\textwidth]{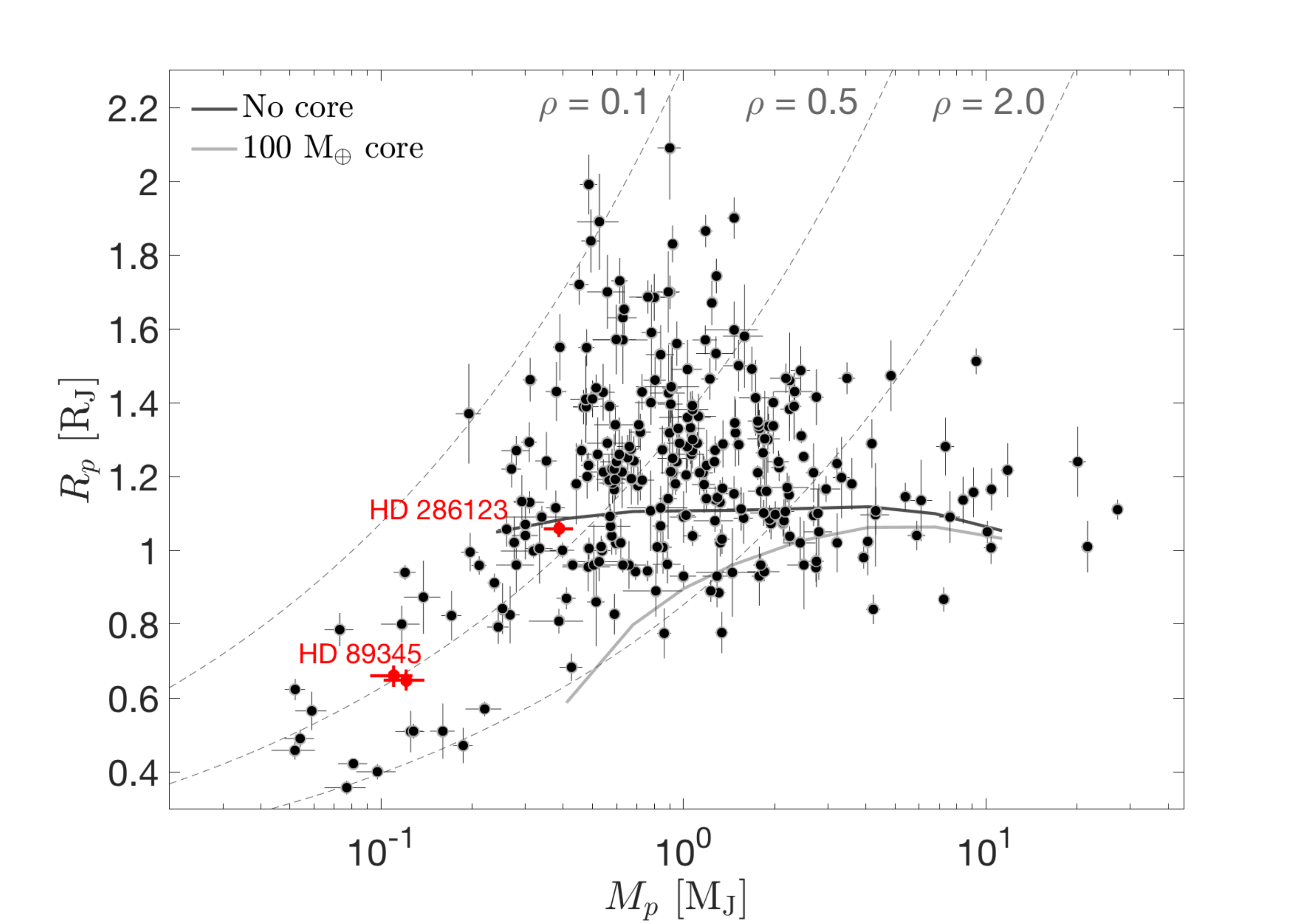}
\caption{Radius-Mass diagram (X-axis in log scale) of transiting planets with measured masses and radii, for planets with $R_p > 0.3\ \Rj$. The two new planets described in this paper are labeled and marked in red. For HD~89345 both solutions are marked in the plot. The thick solid and dashed lines show radius-mass models from \cite{fortney07} for gas giants with no solid core (thick solid black line) and a large core of 100 \Mearth\ (thick solid gray line). Also plotted are three equal density lines (dashed thin gray lines) with mean densities of $<\!\rho\!>$ = 0.1, 0.5, and 2.0 g cm$^{-3}$. 
Data retrieved from the NASA Exoplanet Archive on April 20, 2018.}
\label{fig:massrad} 
\end{figure*}

\begin{figure*}[hbt]
\centering
\includegraphics[width=0.75\textwidth]{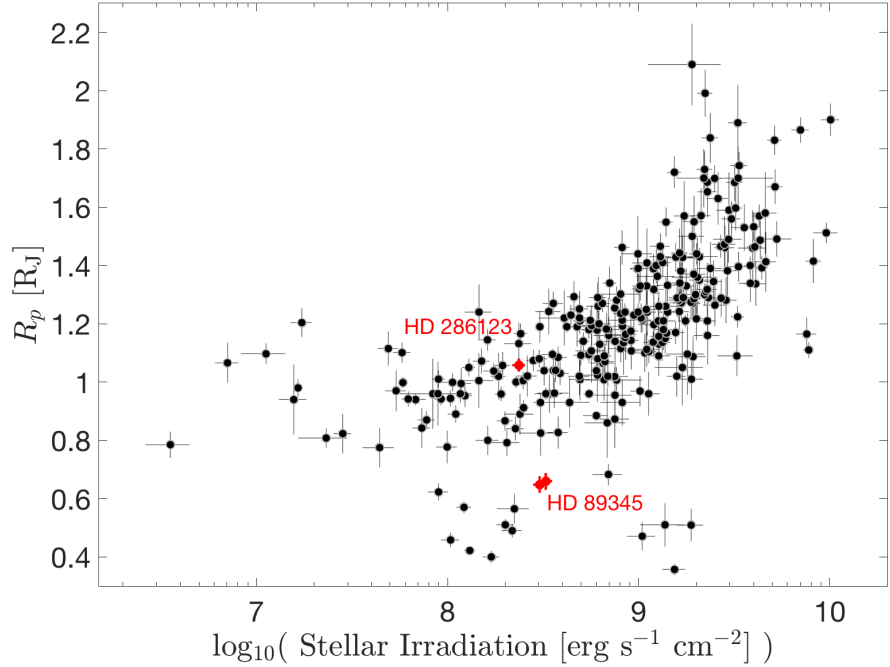}
\caption{Planet radius vs.~stellar irradiation at the planets' orbits for transiting planets with measured mass and radius, for planets with $R_p > 0.3\ \Rj$. The two new planets described in this paper are labeled and marked in red. For HD~89345 both solutions are marked in the plot.
Data retrieved from the NASA Exoplanet Archive on April 20, 2018.}
\label{fig:irrad} 
\end{figure*}

\pltwo and \plthree are gas giants on roughly 11-day period orbits around moderately evolved stars. At ages of roughly 4-7 Gyr, the host stars are near the end of or already leaving the main sequence. The time-averaged incident flux on the planets are given in Table~\ref{tbl:Parameters} as $(3.03 \pm 0.22) \times 10^8$ erg s$^{-1}$ cm$^{-2}$ (solution 1) or $(2.80 \pm 0.20) \times 10^8$ erg s$^{-1}$ cm$^{-2}$ (solution 2) for \pltwo and $(2.14 \pm 0.11) \times 10^8$ erg s$^{-1}$ cm$^{-2}$ for \plthree, all just above the observed radius inflation threshold found by \citet{millerfortney} and \citet{demoryseager}. Yet, when shown in a mass-radius diagram (Fig.~\ref{fig:massrad}) alongside other planets with measured masses and radii, neither appears unusually large for its mass. The same conclusion can be drawn from Fig.~\ref{fig:irrad}, where the radii of the two planets are compared with those of other planets at similar irradiation levels. Thus, despite being slightly above the critical insolation required for radius inflation, neither planet is significantly inflated.

\begin{figure*}[hbt]
\centering
\includegraphics[width=0.32\textwidth, trim={0.5cm 0cm 0.5cm 0cm}, clip]{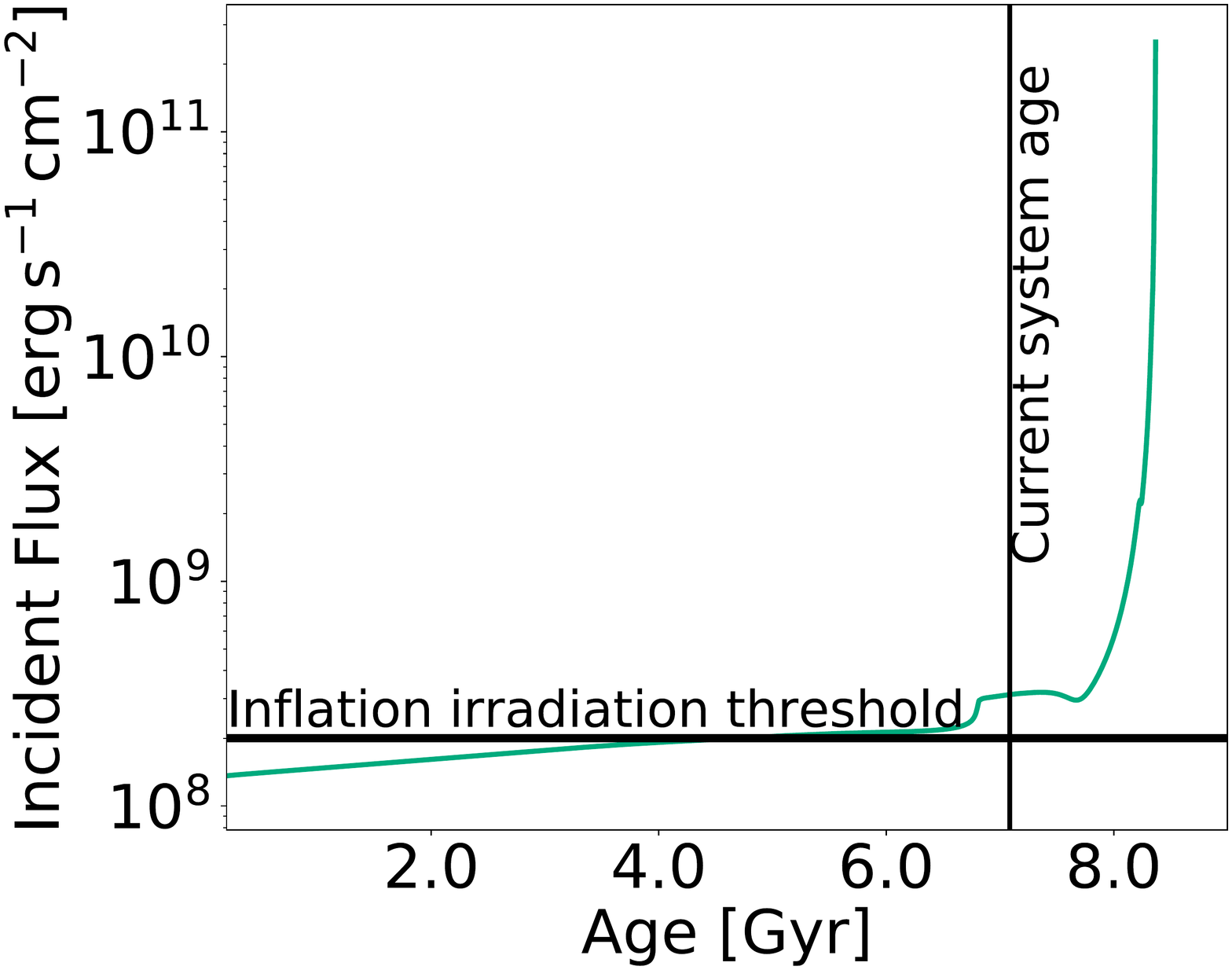} \hfill
\includegraphics[width=0.32\textwidth, trim={0.5cm 0cm 0.5cm 0cm}, clip]{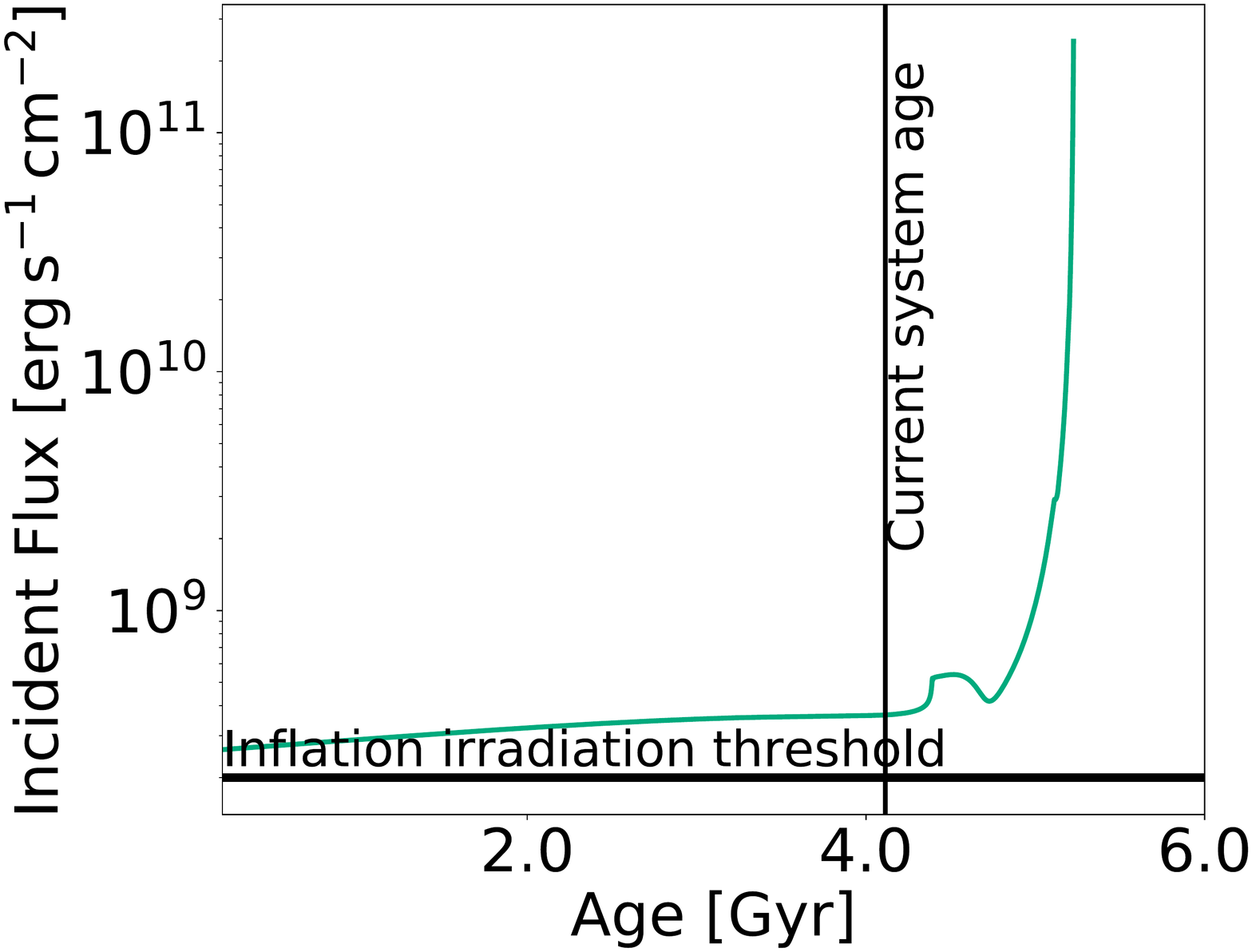} \hfill
\includegraphics[width=0.32\textwidth, trim={0.5cm 0cm 0.5cm 0cm}, clip]{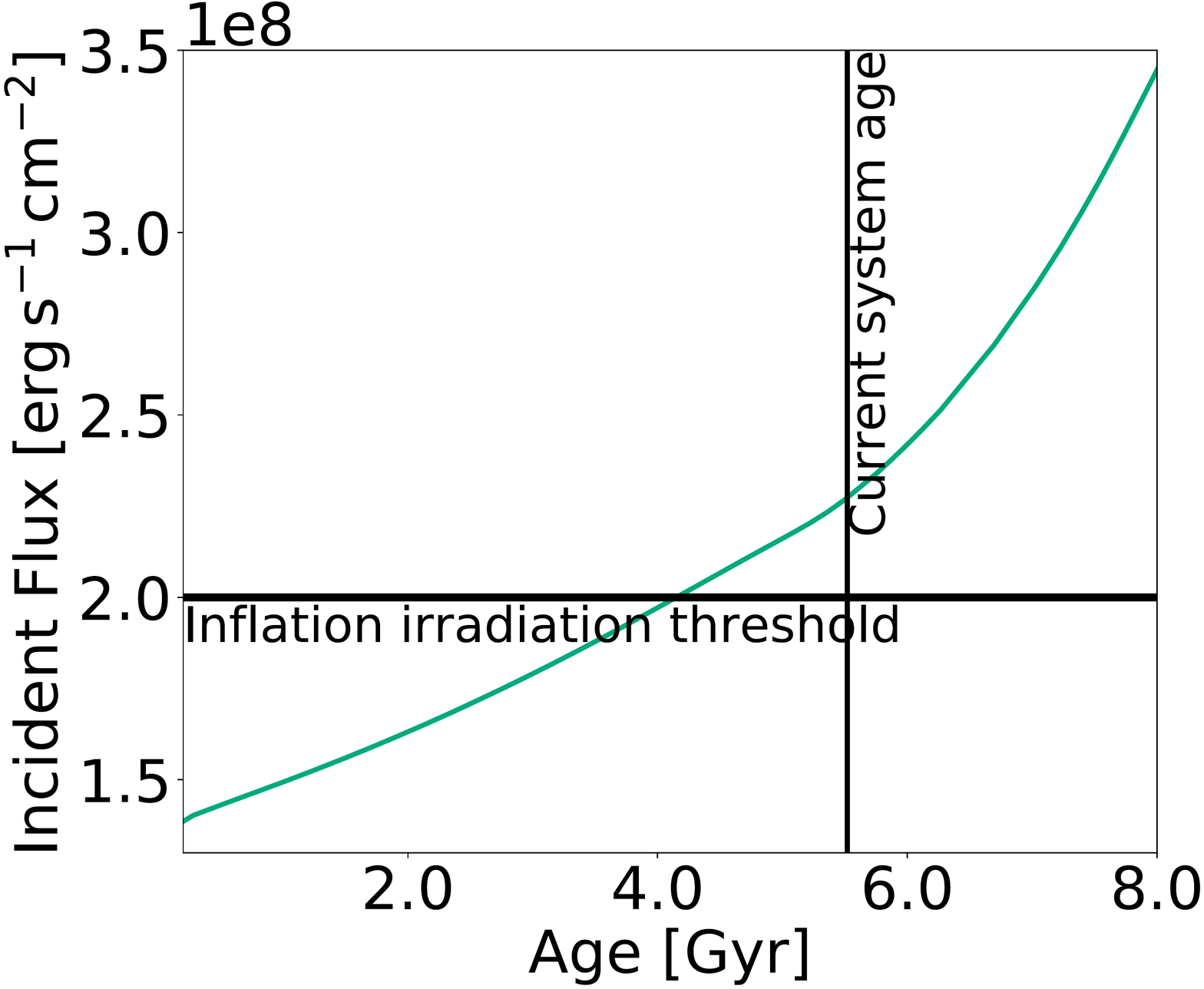}
\caption{Models of the orbital evolution of \pltwo solution 1 (left),  \pltwo solution 2 (right) and \plthree (right) for tidal quality factors $Q'_\star \sim 10^5$ and $Q'_\mathrm{planet} \sim 10^6$. The horizontal line represents the threshold value of $2\times10^8\,\mathrm{erg\  s^{-1} \ cm^{-2}}$ for radius inflation from \citet{millerfortney} and \citet{demoryseager}. The vertical lines show the current ages of the systems.}
\label{fig:evhist} 
\end{figure*}

To further examine the irradiation history of these two planets, we estimate the change in stellar irradiation over time using MIST evolutionary tracks \citep{dotter16, choi16} interpolated to stellar masses and metallicities derived in Section~\ref{sec:exofast}. Fig.~\ref{fig:evhist} shows
the irradiation history of both planets as their host stars evolve. 
We conclude that for both planets, the orbit-averaged incident flux has been within a factor of two of the empirical critical value of $\sim 2\times10^8\,\mathrm{erg\  s^{-1} \ cm^{-2}}$ at least as far back as the zero-age
main sequence phase of the host stars. 

The above calculation ignores possible evolution in the orbits of the planets.
This is justified in the absence of other bodies in the systems, since the only
other mechanism for orbital evolution is tidal decay after the disk disappears, and for both systems the
timescales of this process are rather long, even assuming efficient dissipation
(tidal quality factors of $Q'_\star \sim 10^5$ and $Q'_\mathrm{planet} \sim 10^6$) and
taking the present day planetary and stellar radii, which must have been smaller
in the past. In particular, using Equations 1 and 2 from
\citet{Jackson_Greenberg_Barnes_2009}, the timescales for the evolution of the
semi-major axis and the orbital eccentricity are approximately

\begin{eqnarray}
    \left(\frac{1}{a} \frac{da}{dt}\right)^{-1} \approx 210\,\mathrm{Gyr \ (sol.1) \ or} \ 290\,\mathrm{Gyr \ (sol.2)}
    \quad \\
    \left(\frac{1}{e} \frac{de}{dt}\right)^{-1} \approx 40\,\mathrm{Gyr \ (sol.1) \ or}\ 50\,\mathrm{Gyr \ (sol.2)}
    \quad & \quad 
\end{eqnarray}
for \pltwo, and 
\begin{eqnarray}
    \left(\frac{1}{a} \frac{da}{dt}\right)^{-1} \approx 85\,\mathrm{Gyr}
    \quad \\
    \left(\frac{1}{e} \frac{de}{dt}\right)^{-1} \approx 20\,\mathrm{Gyr}
    \quad & \quad 
\end{eqnarray}
for \plthree. Using $Q'_\mathrm{planet} \sim 10^5$ for \plthree results in an eccentricity decay timescale
of just $3\,\mathrm{Gyr}$, which conflicts with the observed
non-zero eccentricity of the system.

The results of our calculation therefore apply to any dissipation less efficient that $Q'_\star \sim 10^5$ and $Q'_\mathrm{planet} \sim 10^6$. In this regime, both planets have been very close to the critical irradiation threshold throughout their lifetimes. \citet{lopezfortney} found that if the inflation mechanism operates by depositing some fraction of a planet's incident irradiation into its deep interior (class I), then a Saturn-mass planet on a 20-day orbit around a 1.5 \Msun star can rapidly inflate to more than 2 Jupiter radii as the host star leaves the main sequence. In contrast, a class II inflation mechanism that operates by delayed cooling should not cause a planet to inflate as its host evolves off the main sequence. We stress that the critical irradiation threshold is not known to better than a factor of two. That the two planets presented here are not inflated shows that if class I mechanisms are indeed responsible for planet inflation, then these planets have not yet reached high enough irradiation levels or have not had time to inflate in response to increasing irradiation. Regardless, they probe the regime where inflation begins to be noticeable, and provide two new additions to the currently very small sample of warm gas giants to test the two theories. Moreover, most existing gas giant inflation studies have focused on Jupiter-mass objects, but these new detections are lower mass and could potentially provide interesting new insight into the physical processes governing inflation.

\subsection{\pltwo and the Transition between Ice Giants and Gas Giants}
\pltwo has a radius 0.8 times that of Saturn and a mass $\sim0.1$ times that of Jupiter. It may therefore be a rare example of a sub-Saturn ($4\ \Rearth < R_p < 8 \ \Rearth$, and $0.02\ M_\mathrm{J} \lesssim M_p \lesssim 0.2\ M_\mathrm{J}$, using the definition of \citet{petigura16}). Apart from \pltwo, there are only $\sim 20$ known sub-Saturns with masses determined to within 50\% accuracy. In the core accretion scenario, rapid accretion of a gaseous envelope is expected to start in this mass regime \citep[e.g.][]{mordasini15}. Sub-Saturns are therefore an important mass regime for studying the transition between ice giants and gas giants. 

Sub-Saturns have no analogs in the Solar System, but may shed light on the formation mechanisms of similar intermediate-mass planets in the Solar System (Uranus and Neptune). It is commonly assumed that ice giants like Uranus and Neptune formed via core accretion. Under this assumption, the accretion rate must be high enough to ensure that enough gas is accreted, but with high accretion rates, such planets would become gas giants the size of Jupiter and Saturn, instead of ice giants \citep[e.g.][]{helledbodenheimer}. To explain the formation of ice giants, core accretion models must prematurely terminate their growth by dispersal of the gaseous disk during envelope contraction \citep{pollack96, dodsonrobinson}.

At a period of 11.8 days, \pltwo is much closer to its host star than the Solar System ice giants. Under the core accretion scenario, at such small radial distances, where the solid surface density is high, planets are even more likely to undergo runaway accretion that turns them into gas giants. It would therefore be interesting to see whether the composition of \pltwo more closely resembles that of ice giants or gas giants. One way to test this is to measure the atmospheric metallicity of the planet through transmission spectroscopy, since the Solar System's ice giants have significantly higher atmospheric metallicities compared to the gas giants \citep{guillotgautier}.

\section*{Acknowledgments}
During the completion of this paper, we became aware of another paper reporting the discovery of a planet orbiting \candthree \citep{brahm18}. During the referee process, another paper \citep{vaneylen18} independently reported the discovery of \pltwo.

We thank Chelsea Huang for helpful discussions on the manuscript. This work made use of the SIMBAD database (operated at CDS,
Strasbourg, France) and NASA's Astrophysics Data System Bibliographic
Services. This research has made use of the NASA Exoplanet Archive, the Exoplanet Follow-up Observing Program (ExoFOP), and
the Infrared Science Archive, which are operated by the California
Institute of Technology, under contract with the National Aeronautics
and Space Administration. This publication makes use of data products from the Wide-field Infrared Survey Explorer, which is a joint project of the University of California, Los Angeles, and the Jet Propulsion Laboratory/California Institute of Technology, funded by the National Aeronautics and Space Administration. The authors
wish to recognize and acknowledge the very significant cultural role
and reverence that the summit of Mauna Kea has always had within the
indigenous Hawaiian community.  We are most fortunate to have the
opportunity to conduct observations from this mountain. 
A portion of this  work was supported by a NASA Keck PI Data Award, administered by the NASA Exoplanet Science Institute. Data presented herein were obtained at the W. M. Keck Observatory from telescope time allocated to the National Aeronautics and Space Administration through the agency's scientific partnership with the California Institute of Technology and the University of California. The Observatory was made possible by the generous financial support of the W. M. Keck Foundation.
This work was performed in part under contract with the California Institute of Technology/Jet Propulsion Laboratory funded by NASA through the Sagan Fellowship Program executed by the NASA Exoplanet Science Institute.  I.J.M.C. acknowledges support from NASA through  K2GO grant 80NSSC18K0308 and from NSF through grant AST-1824644. M.B. acknowledges support from the North Carolina Space Grant Consortium. Work performed by J.E.R. was supported by the Harvard Future Faculty Leaders Postdoctoral fellowship.

{\it Facility:} \facility{Kepler}, \facility{K2}, \facility{Keck-I (HIRES)}, \facility{Keck-II (NIRC2)}, \facility{Palomar:Hale (PALM-3000/PHARO)}, \facility{FLWO: 1.5 m (TRES)}, \facility{APF}


\end{document}